\documentclass[fleqn,usenatbib]{mnras}
\usepackage{newtxtext,newtxmath}
\usepackage[T1]{fontenc}
\DeclareRobustCommand{\VAN}[3]{#2}
\let\VANthebibliography\thebibliography
\def\thebibliography{\DeclareRobustCommand{\VAN}[3]{##3}\VANthebibliography}

\usepackage{graphicx}	% Including figure files
\usepackage{amsmath}	% Advanced maths commands

\defcitealias{Levan25}{L25}

\title[GRB 250702B as a WD-IMBH TDE]{Can GRB 250702B be explained as the tidal disruption of a white dwarf by an intermediate mass black hole? Yes.}

\author[R. A. J. Eyles-Ferris et al.]{
Rob A. J. Eyles-Ferris,$^{1}$\thanks{E-mail: raje1@leicester.ac.uk}
Andrew King,$^{1,2,3}$
Rhaana L. C. Starling,$^{1}$
Peter G. Jonker,$^{4}$
Andrew J. Levan,$^{4}$
\newauthor
Antonio Martin-Carrillo,$^{5}$
Tanmoy Laskar,$^{6,4}$
Jillian C. Rastinejad,$^{7,8}$ 
Nikhil Sarin,$^{9,10}$
Nial R. Tanvir,$^{1}$
\newauthor
Benjamin P. Gompertz,$^{11,12}$
Nusrin Habeeb,$^{1}$
Paul T. O'Brien$^{1}$
and Massimiliano De Pasquale$^{13}$\\
$^{1}$School of Physics and Astronomy, University of Leicester, University Road, Leicester, LE1 7RH, UK\\
$^{2}$Astronomical Institute Anton Pannekoek, University of Amsterdam, Science Park 904, NL-1098 XH Amsterdam, The Netherlands\\
$^{3}$Leiden Observatory, Leiden University, Niels Bohrweg 2, NL-2333 CA Leiden, the Netherlands\\
$^{4}$Department of Astrophysics/IMAPP, Radboud University, PO Box 9010, 6500 GL Nijmegen, The Netherlands\\
$^{5}$School of Physics and Centre for Space Research, University College Dublin, Belfield, Dublin 4, Ireland\\
$^{6}$Department of Physics \& Astronomy, University of Utah, Salt Lake City, UT 84112, USA\\
$^{7}$NHFP Einstein Fellow\\
$^{8}$Department of Astronomy, University of Maryland, College Park, MD 20742, USA\\
$^{9}$Kavli Institute for Cosmology, University of Cambridge, Madingley Road, CB3 0HA, UK\\
$^{10}$Institute of Astronomy, University of Cambridge, Madingley Road, CB3 0HA, UK\\
$^{11}$School of Physics and Astronomy, University of Birmingham, Birmingham B15 2TT, UK\\
$^{12}$Institute for Gravitational Wave Astronomy, University of Birmingham, Birmingham B15 2TT, UK\\
$^{13}$MIFT Department, University of Messina, Via F. S. D'Alcontres 31, 98166, Messina, Italy
}

\date{Accepted XXX. Received YYY; in original form ZZZ}

\pubyear{\the\year{}}

\begin{document}
\label{firstpage}
\pagerange{\pageref{firstpage}--\pageref{lastpage}}
\maketitle

\begin{abstract}
GRB 250702B is a unique astrophysical transient characterised by its nature as a repeating gamma-ray trigger. Its properties include possible periodicity in its gamma-ray light curve, an X-ray counterpart that rose prior to the gamma-ray outbursts and faded quickly, and radio and infrared counterparts. These features are difficult to reconcile with most models of high energy transients but we show that they are compatible with a white dwarf bound to an intermediate mass black hole that is tidally stripped over multiple pericentre passages before being fully disrupted. In this model, accretion onto the black hole powers a mildly relativistic jet that produces the X-rays through internal processes and the infrared and radio counterparts through thermal emission and external shocks respectively but is unable to produce the gamma-ray emission on its own. We find that chaotic debris streams from the multiple stripping episodes can collide with a period roughly the same as the orbital period of the star. These shocks produce X-ray photons that are upscattered by the jet to produce the observed MeV gamma-ray emission. Future analysis of the jet properties will allow us to place firmer constraints on our model. 
\end{abstract}

\begin{keywords}
gamma-ray burst: individual: GRB 250702B -- tidal disruption events -- accretion -- jets
\end{keywords}

%%%%%%%%%%%%%%%%%%%%%%%%%%%%%%%%%%%%%%%%%%%%%%%%%%

%%%%%%%%%%%%%%%%% BODY OF PAPER %%%%%%%%%%%%%%%%%%

\section{Introduction}

Gamma-ray bursts (GRBs) are fast-evolving sources of luminous gamma-rays typically powered by either the collapse of a massive star or the merger of two compact neutron stars \citep[e.g.][]{Woosley06,Abbott17}. GRBs typically last seconds to minutes, with the hours long ultra-long GRBs (ulGRB) population representing an outlier \citep[e.g.][]{Levan14}, and the nature of their progenitor systems means they are singular events. The striking feature of the enigmatic gamma-ray transient GRB 250702B, however, is its repeating nature, triggering the \textit{Fermi} mission's Gamma-ray Burst Monitor (GBM) instrument three separate times over several hours \citep{GCN40883,GCN40886,GCN40890,GCN40891}. This very high energy emission was detected by several other missions \citep{GCN40903,GCN40910,GCN40914,GCN40923} and visual inspection of the \textit{Fermi} light curves show the individual triggers\footnote{Available from \url{https://heasarc.gsfc.nasa.gov/FTP/fermi/data/gbm/triggers/2025/} under trigger IDs \texttt{bn250702548}, \texttt{bn250702581} and \texttt{bn250702682}.} to be rapidly variable on timescales of $\sim$seconds. Reanalysis of previous observations also revealed an X-ray counterpart detected by the Wide-field X-ray Telescope (WXT) on board \textit{Einstein Probe} (hereafter \textit{EP}) a day prior to the gamma-ray triggers \citep{GCN40906}. This behaviour is difficult to explain through the standard fireball model of collapsar driven GRBs and the timescales observed are significantly longer than in the similarly complex ulGRB sample \citep{Levan14}.

The localisation offered by \textit{EP} allowed further follow-up of the X-ray counterpart from the \textit{Neil Gehrels Swift Observatory} (hereafter \textit{Swift}), \textit{EP}, NuSTAR and the \textit{Chandra X-ray Observatory} (hereafter \textit{Chandra}) \citep{GCN40917,GCN40919,GCN41014,GCN41309,GCN41767}. The X-ray counterpart was found to fade rapidly as a power law with temporal index $\alpha\sim1.9$. A radio counterpart was also identified and has been extensively monitored \citep{GCN40979,GCN40985,GCN41053,GCN41054,GCN41059,GCN41061,GCN41145,GCN41147,GCN41215,Levan25}.

At optical wavelengths, follow-up was hindered by significant extinction in the Milky Way and the host. However, Very Large Telescope (VLT) observations were able to identify a fast fading near-infrared (NIR) counterpart \citep{GCN40924,GCN40961,Levan25}. \textit{Hubble Space Telescope} observations conclusively showed the source to be associated with a galaxy. These observations also provided evidence of significant dust lanes which contribute to the extensive extinction \citep{GCN41096,Levan25}.

\citet{Levan25} presented early analysis of GRB 250702B and in particular, showed evidence that the multiple gamma-ray triggers are consistent with a period of 2825 s in the observer frame \citep{Levan25}. Assuming a relatively low redshift of $z\lesssim0.3$, they also showed the X-ray, NIR and optical can be attributed to afterglow emission from a GRB or could be linked to a white dwarf (WD) being tidally disrupted and accreted by an intermediate mass black hole (IMBH). These tidal disruption events (TDEs) arise when the WD passes within the tidal radius, $r_t=R_{\rm WD}\left(M_\bullet / M_{\rm WD}\right)^{1/3}$, of the IMBH. The tidal forces exerted by the IMBH overcome the WD's self-gravity and it is torn apart \citep[see][for a review]{Maguire20}. A significant fraction of the debris remains bound to the IMBH and is accreted to produce a luminous transient. WD-IMBH TDEs have previously been suggested as the progenitor system for ulGRBs \citep[][]{Levan14} as the highly super-Eddington accretion rates can offer suitable conditions for a jet to be launched. The fast X-ray transient XRT 000519 \citep{Jonker13} may also be an example of such a system and has several remarkable properties and similarities to GRB 250702B, particularly an X-ray precursor and similar evidence of periodicity in its light curve. However, without a redshift for XRT 000519, a direct comparison is difficult.

The host of GRB 250702B has now been observed by the \textit{James Webb Space Telescope} \citep[\textit{JWST,}][]{Gompertz25} which has identified the redshift of $z=1.036\pm0.004$ and allowed the energetics of the transient to be constrained to an isotropic equivalent energy of order $2\times10^{54}$ erg. However, there is no evidence of transient light and supernovae, in particular, are limited to only the faint outliers of typical GRB-supernovae. The availability of the redshift has allowed several models of GRB 250702B's behaviour to be assessed although its nature remains ambiguous \citep[see e.g.][for an overview]{Carney25,OConnor25}. These include a micro-TDE where a main sequence star is disrupted by a stellar mass black hole \citep{Beniamini25,OConnor25} and a merger between a stellar mass black hole \citep{Neights25} while a WD-IMBH TDE remains a possibility \citep{Oganesyan25,Li25}.

In this work, we reevaluate whether GRB 250702B can indeed be explained by a WD-IMBH TDE armed with more knowledge of the long term evolution and, in particular, the redshift and associated energetics. A similar model is examined in \citet{Li25}, although the assumptions made there differ significantly in some areas as briefly discussed in Section \ref{sec:tidal_stripping}. In Section \ref{sec:xray}, we analyse X-ray observations from both \textit{Swift} and \textit{Chandra}. We present and discuss our model of a repeatedly stripped WD that is finally disrupted to produce GRB 250702B and its counterparts in Section \ref{sec:model} and finally summarise our conclusions in Section \ref{sec:conclusions}. Throughout this Letter, we assume a Planck cosmology \citep{Planck20}.

\section{X-ray data analysis}
\label{sec:xray}

Thanks to the redshift identified by \citet{Gompertz25}, we are able to place firmer constraints on the X-ray properties and therefore the overall nature of GRB 250702B. Here, we reanalyse the data obtained by \textit{Swift}-XRT and present our analysis of two \textit{Chandra} DDT obervations. Errors are given to 90\% confidence unless otherwise stated.

\subsection{\textit{Swift}-XRT}

We retrieved the spectral data and count rate light curve of GRB 250702B from the UK \textit{Swift} Science Data Centre \citep[UKSSDC\footnote{\url{https://www.swift.ac.uk/index.php}},][]{Evans09}. These data included all photons detected by \textit{Swift}-XRT and we fitted them with an absorbed power law model using \textsc{XSpec v12.13.1} \citep{Arnaud96}. We used two absorption  components, a Galactic absorber with fixed $N_{H,~{\rm gal}} = 3.6\times10^{21}$ cm$^{-2}$ \citep{Willingale13} and a second absorber at the redshift of GRB 250702B to represent host absorption $N_{H,~{\rm host}}$. We find the spectrum is broadly typical for a transient X-ray source with photon index $\Gamma=1.76\pm0.21$ and infer a high host column density $N_{H,~{\rm host}} = 5.1 \pm 1.6 \times 10^{22}$ cm$^{-2}$, consistent with the extensive dust found along the line of sight \citep[][]{Levan25,Gompertz25}.

We also checked for spectral evolution by fitting the data from the first epoch of \textit{Swift}-XRT observations separately. Using the same model, we find $\Gamma=1.56\pm0.21$ and $N_{H,~{\rm host}} = 3.86 \pm 1.49 \times 10^{22}$ cm$^{-2}$, consistent with the spectrum reported for \textit{EP} Follow-up X-ray Telescope (FXT) observations at a contemporary time \citep{GCN40917}. We therefore find no significant evidence for spectral evolution and use the model inferred from our total spectrum fit to convert the count rate light curve to 0.3 - 10 keV luminosities\footnote{We also note that the counts to luminosity conversion varies by a factor $\sim1.01$ between these models and will therefore have negligible impact on our subsequent analysis.}.

\subsection{\textit{Chandra} ACIS-S}

In addition to \textit{Swift}, GRB 250702B was also monitored through three \textit{Chandra} DDT observations of which two are publicly available \citep{GCN41309,GCN41767}, however, we re-examine these data to account for the now known redshift. Both of these observations were performed using the Advanced CCD Imaging Spectrometer (ACIS-S) instrument. We acquired these data from the \textit{Chandra} Data Archive and used the \textsc{Chandra Interactive Analysis of Observations (CIAO) v4.17.0} package \citep{Fruscione06} with \textsc{CALDB v4.12.2} to analyse each observation.

ObsID 31011 was taken $\sim38$ days after the first gamma-ray trigger and lasted for 27.71 ks. We used the \texttt{specextract} CIAO script to extract the spectrum and the accompanying response files but due to a gain calibration issue (private communication), we do not use these for spectral analysis. Instead, we assume the same spectrum as the \textit{Swift}-XRT data to convert from our measured count rate of $9.75 \pm 1.94 \times 10^{-4}$ count s$^{-1}$ to the 0.3-10 keV luminosity, similarly to \citet{GCN41309}.

By the time \textit{Chandra} observed ObsID 31468 at ~65 days post trigger, GRB 250702B's counterpart had faded significantly. Instead of attempting to extract a spectrum, we therefore used the \texttt{srcflux} CIAO script to measure a count rate of $1.12^{+1.27}_{-0.74} \times 10^{-4}$ count s$^{-1}$. Again, we use the same spectrum as the \textit{Swift}-XRT observations to convert this to a 0.3-10 keV luminosity.

\begin{figure}
	\includegraphics[width=\columnwidth]{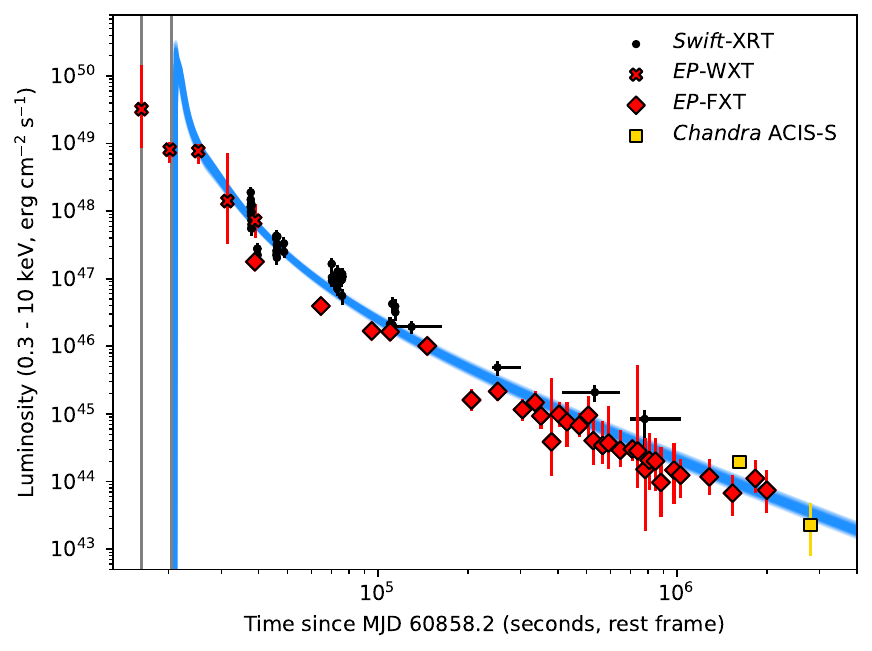}
    \caption{The long term X-ray light curve of GRB 250702B compared to all 177 1-$\sigma$ traces from our MCMC fit using a WD-IMBH model (blue solid lines). The vertical grey lines mark the times of the Fermi triggers. Note the first plotted WXT point is excluded from our fitting.}
    \label{fig:xray_lc}
\end{figure}

We show the complete X-ray light curve, including the data from \textit{EP} observations described by \citet{Li25}, in the context of a WD-IMBH TDE in Figure \ref{fig:xray_lc}.

\section{A WD-IMBH TDE model of GRB 250702B}
\label{sec:model}

Returning to the observed properties of this event, there are five key features that any attempt to explain GRB 250702B must address:
\begin{enumerate}
    \item \textit{Precursor X-ray emission from $\sim$few 10 ks prior to the gamma-ray flares.}
    \item \textit{High peak X-ray luminosity that quickly fades as $\sim t^{-1.9}$.}
    \item \textit{A fast decaying NIR counterpart and long lived radio counterpart.}
    \item \textit{Gamma-ray flares peaking at hundreds of keV and stretching to MeV energies with fast ($\sim$seconds timescale) variability.}
    \item \textit{Near constant intervals between the gamma-ray flare times.}
\end{enumerate}

We now examine GRB 250702B further in the context of a WD being disrupted by an IMBH and whether these features can be produced by such an event. We make several key assumptions - the periodicity investigated by \citet{Levan25} is both real and related to the orbital period of the WD\footnote{This implies a significantly shorter orbital period than assumed in \citet{Li25}.}, and that the conditions, such as the properties of the magnetic field near the IMBH, are sufficient to launch a relativistic jet. Importantly, as shown by \citet{Levan25}, the orbital period eliminates both supermassive black holes and main sequence stars as components of the progenitor system.

\subsection{Repeated tidal stripping and accretion of a white dwarf}
\label{sec:tidal_stripping}

A key property of WDs is the inverted relationship between their mass and their radius - lower mass WDs have larger radii and vice versa \citep[e.g.][]{Ambrosino20}. This particular property offers an explanation for the precursor X-ray emission observed in GRB 250702B. Analysis of extreme mass ratio inspirals \citep[EMRIs,][]{AmaroSeoane18} show that while gravitational wave emission does result in a decay in the orbital semi-major axis, this effect is slow and therefore the WD's orbit is, at least initially, relatively stable \citep[e.g.][]{Peters64}.

However, as the WD undergoes a pericentre passage, while its core may be outside the tidal radius $r_t$, it is possible that the outer layers may be sufficiently weakly bound that a small amount of mass is stripped and accreted by the IMBH. Assuming the WD is able to expand quickly enough to quench any thermonuclear burning which can be induced through compression of the WD in deep encounters \citep{Rosswog08,Rosswog09,Macleod16}, i.e. $t_{\rm dyn} < t_{\rm burning}$, the now lower mass WD will therefore expand in radius. There may also be additional expansion due to tidal distortion although fully quantifying this effect is beyond the scope of this work \citep[e.g.][]{Maguire20}. This stripping will be repeated on the subsequent pericentre passages and is a runaway effect - the WD will have an incrementally larger radius on each passage and therefore more matter will be stripped - until the WD has lost sufficient mass that it will lie within $r_t$ and be fully disrupted. This model is similar to that proposed by \citet[][see also \citet{King20} and \citet{Chen22}]{King22} to explain quasi-periodic eruptions (QPEs) from galactic nuclei. In fact, it is likely that the early tidal stripping events may resemble QPEs as the accretion rate will decay significantly quicker than a later full disruption.

\citet{King22} shows that for most systems, the mass transfer is stable and leads to an expanding orbit to conserve angular momentum. However, in this case, the orbital period is significantly shorter than the viscous timescale (from Equation 25 of \citet{King22} and assuming a modest $H/R\sim0.1$, we find it to be of order several thousand seconds, significantly shorter than that assumed by \citet{Li25}) which means that an accretion disk will not be efficiently formed before the WD returns to pericentre. In this case, then, angular momentum cannot be transferred back to the WD and the orbital semi-major axis does not expand. The short orbital period is crucial in this case and is the direct cause of the runaway. It is important to also note that although the stripping episodes result in fallback rates that evolve quicker than that from the final disruption, they can achieve comparable peak accretion rates \citep{Nixon21}.

Due to the repeated stripping events, shocks between the different debris streams are still likely to induce the formation of an accretion flow at early times \citep{Rosswog09,Bonnerot21b}\footnote{However, the transfer of angular momentum from a particular debris stream still cannot occur before the next pericentric passage due to the long viscous timescale compared to the orbital period. This means the WD will maintain its orbit and the runaway stripping episodes will continue}. This means that material from later passages will eventually be accreted and, due to the low mass of the IMBH involved, will result in highly super-Eddington accretion. In fact, assuming a $10^4~M_\odot$ IMBH, the Eddington luminosity is of order $10^{42}$ erg s$^{-1}$. From Figure \ref{fig:xray_lc}, the vast majority of the stripped material will be accreted over $\lesssim10^6$ s so stripping and accreting as little $10^{-5}~M_\odot$ will result in super-Eddington accretion. The timescale over which the jet launch, and therefore bright X-ray emission becomes observable, occurs is unclear and due to the complex nature of the WD's radius evolution and the amount of matter stripped, we leave this to a future study. However, the jet may only be launched only a few tens of passages prior to the full disruption of the WD as disordered magnetic fields may require the accretion rate to be several orders of magnitude super-Eddington.

The orbit of the WD will naturally differ from that expected in other TDE systems. Typically a parabolic ($e = 1$) orbit is assumed but this results in a single encounter as the star is disrupted. Instead, we require the WD to be bound to the IMBH to ensure continuous orbits (i.e. $e<1$). Achieving this likely requires Hills capture where the WD is initially in a binary system that is disrupted by the IMBH leaving only the WD bound \citep{Hills88} and has been suggested in the case of other repeating transients \citep[e.g.][]{King22,Cufari22}. We discuss this further in Section \ref{sec:formation}.

\subsection{A wide angle, low Lorentz factor jet}

This model of the early accretion and a jet launched through super-Eddington accretion can easily account for the precursor X-ray emission observed by \textit{EP}. In this case, we assume the X-rays originate from some internal process in the jet and this emission dominates over any synchrotron from external shocks. We therefore assume a constant radiative and accretion efficiency in the X-ray band, $\epsilon_X$, and jet (half-)opening angle, $\theta_j$. The X-ray luminosity should therefore be proportional to the total fallback rate. Our model predicts a stepped or multi-peaked rise with a period approximately equal to the 1390 s (converted to rest frame) identified by \citet{Levan25} and remarkably similar to the behaviour exhibited by XRT 000519 \citep{Jonker13} and predicted in \citet{Macleod14}. We note that this is not readily apparent in the precursor light curve as presented by \citet{Li25} but is not necessarily ruled out and further investigation of those data may be indicative of our predicted behaviour. We also note, however, that we have only very weak constraints on the amount of material stripped, and this periodicity could be smoothed out somewhat over the tens of contributing pericentric passages. We therefore do not include the precursor emission our fitting below. Due to the low mass of the IMBH, and in direct contrast to the prediction from \citet{Levan25} which assumed a low redshift event, the accretion rate is likely to stay highly super-Eddington for months\footnote{The redshift and therefore luminosity assumed in \citet{Levan25} were much smaller and therefore significantly closer to the Eddington luminosity,}. The same jet internal processes should therefore dominate the X-ray light curve over the first days to months while at late times, there may be a sudden drop as the accretion rate drops below Eddington as observed in other relativistic TDEs\footnote{It is difficult to say whether this is being seen in the last \textit{Chandra} observation as the difference with the expected power law decay is low significance.}. However, as this is not apparent in the observed light curve, we assume we are in the jet dominated phase throughout.

\begin{figure}
	\includegraphics[width=\columnwidth]{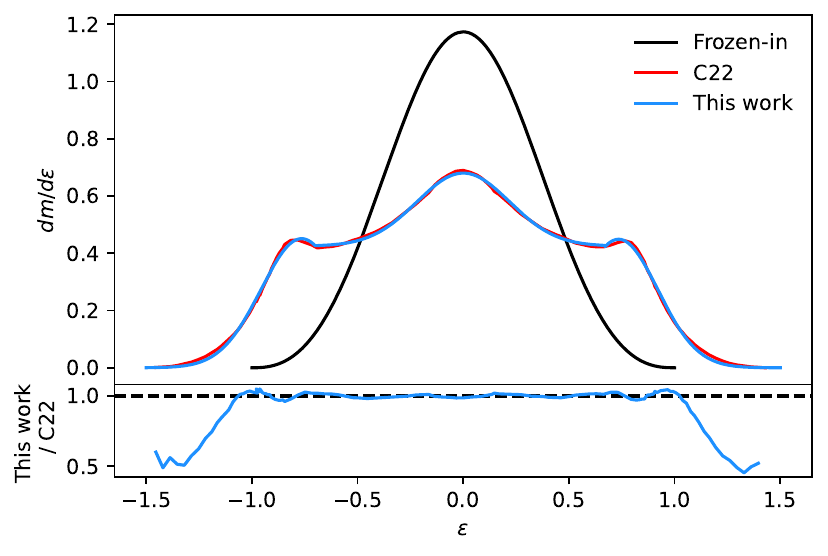}
    \caption{\textbf{Top:} the differential mass per unit specific orbital energy distribution of a disrupted WD derived using the frozen-in approximation (black), numerical simulations \citep[][red, C22]{Cufari22a} and our semi-analytical approximation (blue). An eccentricity of $e=1$ is assumed here. \textbf{Bottom:} the ratio between our semi-analytical approximation and the numerical result of \citep{Cufari22a}.}
    \label{fig:saa_comp}
\end{figure}

Instead of the `frozen-in' approximation, we use a semi-analytical approximation for the fallback rate which partially takes into account the effects of tidal distortion. To briefly summarise, the frozen-in approximation assumes the differential mass per unit specific orbital energy distribution of the debris stream is fixed from the time of disruption. However, numerical simulations have shown this not to be the case due to the tidal distortion of the star at the time of its disruption and the self-gravity of the stream. We therefore use a semi-analytical approximation for the distribution of differential mass per unit specific orbital energy given by
\begin{equation}
    dm / d\varepsilon = \begin{cases}
        A_1 e^{-\frac{(\varepsilon-\varepsilon_{0,1})^2}{2\sigma_1^2}}, & {\rm for}~\varepsilon < \varepsilon_{c,1}, \\
         k_c + A_c e^{-\frac{(\varepsilon-\varepsilon_{0,c})^2}{2\sigma_c^2}}, & {\rm for}~\varepsilon_{c,1} \leq \varepsilon \leq \varepsilon_{c,2}, \\
         A_2 e^{-\frac{(\varepsilon-\varepsilon_{0,2})^2}{2\sigma_2^2}}, & {\rm for}~\varepsilon > \varepsilon_{c,2},
    \end{cases}
    \label{eq:saa_dist}
\end{equation}
where $A_x$, $\varepsilon_{0,x}$ and $\sigma_x$ are the amplitudes, centres and widths of three Gaussians and $k_c$ is an additional constant added to the central Gaussian. These represent the `wings' and core of the star. We fit this model to the numerical results of \citep{Cufari22a} using a least squares method and present our results in Figure \ref{fig:saa_comp}. Figure \ref{fig:saa_comp} shows that our semi-analytical approximation accurately recreates the structure of the numerical result and is a significant improvement on the frozen-in approximation. While the differential masses at the extremes of the specific orbital energy are underestimated, the effect of this is negligible over the timescales we focus on. The more circular WD orbit required here will shift this distribution towards the negative with an offset of $- (1-e) / (1 - e_{\rm crit}^-)$ where $e_{\rm crit}^- = 1 - 2\beta\left(\frac{M_\bullet}{M_{\rm WD}}\right)^{-1/3}$ is the critical eccentricity \citep{Hayasaki18}, $\beta = \frac{r_t}{r_p}$ is the impact parameter and $r_p$ is the pericentric radius. We assume $\beta=1$ throughout. At a given time, $t$, material with a specific orbital energy will return to the IMBH,
\begin{equation}
    \varepsilon(t) = - \frac{R_{\rm WD}}{2} \left(\frac{M_\bullet}{M_{\rm WD}}\right)^{2/3} \left(\frac{\sqrt{G M_\bullet}}{2\pi}\right)^{-2/3} t^{-2/3}
    \label{eq:epsilon_t}
\end{equation}
and we can then simply combine Equations \ref{eq:saa_dist} and \ref{eq:epsilon_t} to derive the fallback rate, $\dot{m}$ at time $t$. As we assume the X-rays are powered by internal processes with constant efficiency, we convert the fallback rate to a luminosity as $L_X(t) = \epsilon_T \dot{m} (t)$ where $\epsilon_T = 2\epsilon_X / (1-\cos \theta_j)$.

We assume the final disruption dominates the long term accretion rate as the earlier stripping episodes are both quicker and will result in significantly less mass being accreted. We fit our model to the long term X-ray decay (from the first post-peak WXT detection as this point coincides with the final \textit{Fermi} trigger i.e. the presumed final disruption\footnote{A reasonable fit with a narrower jet and slightly higher black hole and WD masses can also be obtained if the peak detection is included.}) using the MCMC sampler \textsc{emcee v3.1.6} \citep{ForemanMackey13} over 5000 iterations with the first 500 discarded as burn-in. In Figure \ref{fig:xray_lc}, we show the 177 traces within 1-$\sigma$ of our final result from the resulting MCMC chain and demonstrate that our model can indeed reproduce the long term evolution of GRB 250702B. We summarise the priors and posterior distributions in Table \ref{tab:emcee_posterior}. However, our model does overpredict the observed peak luminosity from \textit{EP} by a factor of $\sim$three to ten. This is not entirely unexpected, however, and suggests that energy is being lost to some other process rather than being radiated in X-rays. In Section \ref{sec:shocks_and_gamma_rays}, we discuss the possibility that this missing energy powers inverse Compton upscattering.

\begin{table}
    \caption{The priors and posterior distributions from our MCMC fit to GRB 250702B's X-ray light curve.}
    \label{tab:emcee_posterior}
    \centering
    \renewcommand{\arraystretch}{1.4}
    \begin{tabular}{ccc}
    \hline
    Parameter & Prior & Posterior \\\hline
    $\log \left(M_{\bullet} / M_\odot\right)$ & $2 <\log \left(M_{\bullet} / M_\odot\right)<5$ & $4.02 \pm 0.63$ \\
    $M_{\rm WD}$ ($M_\odot$) & $0.2 < M_{\rm WD} < 1.4$ & $0.45 \pm 0.20$ \\
    $e$ & $\leq 1$ & $0.923 \pm 0.027$ \\
    $\epsilon_T$ & $0 < \epsilon_T < 1$ &$0.50 \pm 0.25$ \\
    \hline
    \end{tabular}
\end{table}

We can also use the assumed radiative efficiency of the model to make inferences about the jet properties. We make the assumption that $0.001\leq\epsilon_X\leq0.01$ (\citet{Crumley16} find $0.01 - 0.1$ while including beaming efficiency) and find that $\theta_j \sim 0.07 - 0.40$ rad. The Lorentz factor is therefore $\Gamma \sim 1 / \theta_j \sim 2.5 - 13.7$. We note that these values are sensitive to our assumed $\epsilon_X$ although our assumption is reasonable. Such a jet is wider than typically observed in long GRBs \citep[e.g.][]{Frail01,Yi17}, but could be an outlier to the population, and comparable to that inferred for the relativistic TDE Swift J1644+57 at early times \citep{Metzger12,Zauderer13} although we note that \citet{Beniamini23} find a significantly larger $\theta_j$ for that TDE. It therefore seems feasible that the inferred jet could form in a WD-IMBH TDE. While the Lorentz factor is small, it will still be sufficient to produce a radio counterpart through external shock induced synchrotron although we do not examine it in detail here nor make a direct comparison with the extensive radio observations of GRB 250702B already performed \citep{GCN40979,GCN40985,GCN41053,GCN41054,GCN41059,GCN41061,GCN41145,GCN41147,GCN41215,Levan25}. By analogy with other relativistic TDEs \citep[e.g.][]{Pasham15,Hammerstein25}, the NIR counterpart will trace thermal emission, possibly from an inflated accretion flow or cooling envelope \citep[e.g.][]{Coughlin14,Dai18,EylesFerris22,Metzger22,Sarin24}. However, this is likely to have a peak bolometric luminosity $\lesssim10^{45}$ erg s$^{-1}$ and follow the X-ray in fading rapidly, as shown by \citet{Levan25}. The non-detection of significant transient light by \citep{Gompertz25} is therefore also consistent with this model.

This model therefore accounts for the X-ray, NIR and radio emission observed in GRB 250702B i.e. key features i -- iii. However, with such a low Lorentz factor and wide opening angle, it is difficult to conceive of how the jet could, on its own, produce the MeV gamma-rays observed in GRB 250702B.

\subsection{Stream induced shocks and gamma-ray emission}
\label{sec:shocks_and_gamma_rays}

There is, however, another possible emission component that we have not yet examined. Following the disruption or stripping of the WD, the disrupted material will be stretched into a long debris stream. This stream will orbit the IMBH and, depending on the exact conditions, may intersect with itself relatively promptly \citep[e.g.][]{Darbha19,Lu20,EylesFerris24} and can influence formation of the accretion flow \citep{Andalman22}. These collisions are highly energetic and release `free' kinetic energy from material that will still be accreted. They can also enhance the accretion rate as the shocked material is rapidly driven onto the IMBH \citep{Andalman22,EylesFerris24}.

In addition to the stream interacting with itself, the earlier tidal stripping of the WD and the resulting debris streams result in additional and, due to our assumed $\beta=1$ curtailing self-intersections, more probable sources of possible interactions. The individual debris streams will behave similarly and in particular, consecutive streams will be almost identical. If, then, enough amount of mass is stripped in a given passage, the debris stream will be long enough that the tip of the debris stream from the subsequent passage will collide with its tail. Such collisions will occur near pericentre and due to the similarity of the debris streams, will be broadly periodic in accordance with the orbital period of the WD. However, whether two consecutive streams collide will also be affected by precession of the debris streams which naturally arises from the fact that the WD's orbit is almost certainly initially unaligned with the spin axis of the IMBH and is unlikely to have become aligned despite the substantial number of orbits since it had become bound \citep[e.g.][]{Stone12,Franchini16,Bonnerot21}\footnote{This misalignment may also lead to jet precession which can further contribute to the periodicity \citep[e.g.][]{Teboul23,An25}.}. This is consistent with the spacing of the three gamma-ray flares - substantial stream interactions occur coincident with the gamma-ray flares but only a small number or only weak collisions may occur between the second two. Alternatively, it is possible that less mass was stripped during the relevant passages and therefore the debris streams were not long enough to intersect. This could be due to the effects of tidal distortion but requires numerical simulations that are beyond the scope of this work to confirm.

The energy released by these collisions is likely to vary somewhat but can be roughly constrained using the formalism of \citet{Darbha19}\footnote{While \citeauthor{Darbha19} concentrate on ultra-deep encounters where a single debris stream promptly intersects with itself, the same physics are applicable to similar debris streams interacting with each other.}. Near the pericentre, the tidal forces exerted by the black hole will compress the streams \citep[e.g.][]{Luminet85} to a radius $\sim R_{\rm WD}$ and the surface density will therefore be comparable to the original WD i.e. $\rho\sim3M_{\rm WD} /4\pi R_{\rm WD}^3$ assuming a constant density within the WD. The kinetic energy of the shocked material is then given by $\rho v_s^2$ where $v_s$ is the shock velocity which is comparable to the pericentric velocity $v_s \sim v_p = \left(GM_\bullet/r_p\right)^{1/2}$. We assume that $r_p = r_t$ and therefore for our derived posteriors, $v_p = 0.23^{+0.21}_{-0.12}~c$. The energy is deposited into an optically thin surface layer with thickness $\Delta R \lesssim (cR_{\rm WD}/\rho \kappa v_p)^{1/2}$, where $\kappa$ is the opacity, and therefore the total energy released thermally is given by 
\begin{equation}
    \Delta E \sim 2\pi R_{\rm WD} l_c \Delta R \rho v_s^2 \sim 2\pi l_c \left(R_{\rm WD}^3  v_p^3 c\rho / \kappa\right)^{1/2}
    \label{eq:shock_energy}
\end{equation}
where $l_c\sim R_{\rm WD}$ is the light crossing time. Assuming a standard opacity of $\kappa = 0.4$ cm$^{2}$ g$^{-1}$, we therefore estimate a luminosity of $4.3^{+9.5}_{-3.2}\times10^{47}$ erg s$^{-1}$ emitted over a time $\sim0.03$ s derived from $l_c$. Assuming a perfect blackbody, the spectrum will peak in the medium to hard X-rays at $15.9^{+7.7}_{-6.0}$ keV.

These collisions may not make a significant contribution to the observed X-ray light curve and despite the relatively high peak energy, are still significantly below the spectrum observed by \textit{Fermi}. They are too faint by over an order of magnitude to trigger \textit{Swift}'s Burst Alert Telescope assuming a 15 - 150 keV sensitivity of $3\times10^{-9}$ erg cm$^{-2}$ s$^{-1}$ \citep[e.g.][]{Lien16}. However, in addition to being observed directly, the emission from the collisions provides a significant field of seed photons that could be upscattered by the jet\footnote{The accretion disk and any disk winds may also provide additional sources of seed photons. However, the luminosities of these components and the energies of the resulting photons are unlikely to make a significant contribution. The jet itself may also provide seed photons for synchrotron self-Compton but due to the low peak energy and small $\Gamma$ this may only be a small contribution at gamma-ray energies. We therefore do not examine them further here.}. If we make the standard approximation of $\nu \sim \Gamma^2\nu_0$ \citep[e.g.][]{Rybicki_and_Lightman,Longair}, the peak in the inverse Compton spectrum should lie in the range $\sim0.6-4.4$ MeV, depending the value assumed for $\epsilon_X$. It is therefore broadly consistent with the $E_p$ identified in the spectral fits of \citet[][see also \citet{Oganesyan25} and \citet{Li25}]{Neights25}. The nature of the seed photon field produced by these shocks can also help explain the highly variable nature of the gamma-rays. As mentioned above, each thermal flash from these collisions will last $\sim0.03$ s and the chaotic nature of the debris streams will therefore cause significant short timescale variability. It is also worth noting that the emission region of these flashes is at $\sim r_t$. For a $M_{\rm WD} = 0.45~M_\odot$, $M_\bullet = 10^4~M_\odot$ system that we assume, this is roughly a lightsecond away from the jet. Variability on timescales of a second in the rest frame is therefore entirely reasonable to expect in the gamma-ray light curve.

A final consideration is the luminosity of the gamma-rays. From \citet{Neights25}, the luminosity of each trigger is a few $10^{51}$ erg s$^{-1}$, significantly higher than the thermal luminosity produced from the stream-stream collisions. The majority of the energy must therefore come from the relativistic electrons within the jet. While this appears to be unachievable from Figure \ref{fig:xray_lc}, the kinetic luminosity of the jet, i.e. the energy available for upscattering the photons, is significantly higher than the X-ray luminosity. From the results of \citet{Wu18}, the kinetic efficiency relative to the fallback rate could be as high as 0.05, a factor of 50 greater than derived for the X-rays. The energy budget is therefore more than sufficient to boost the shock photons and produce the gamma-rays. We note also that this loss of energy to this process is entirely consistent with our model's overprediction of the \textit{EP}-WXT observed luminosity. In fact, even if a smaller kinetic efficiency of only 0.005 is assumed, the converted excess X-ray energy ($\gtrsim10^{54}$ erg) is comparable to the total energy emitted as gamma-rays across all three \textit{Fermi}-GBM triggers ($\sim2\times10^{54}$ erg). This stream collision-upscattering origin therefore fully accounts for the gamma-rays and explains features iv and v.

\subsection{Formation of the required system}
\label{sec:formation}

While the system we have described can produce the electromagnetic signal observed in GRB 250702B, we should briefly consider the likelihood of it arising. As mentioned in Section \ref{sec:tidal_stripping}, the extremely bound nature of the WD's orbit may be rare. From \citet{Hills88} and \citet{Cufari22}, the analytically predicted eccentricity of a bound star following the Hills mechanism is given by
\begin{equation}
    e \simeq 1 - \frac{2}{\beta}\left(\frac{M_\bullet}{M_{\rm WD}}\right)^{-1/3},
    \label{eq:hills_eccentricity}
\end{equation}
where $\beta = r_{t,{\rm binary}} / r_p$ and $M_{\rm WD}$ is the mass of the primary which we assume to be our WD. For our MCMC chain, assuming $\beta=1$, Equation \ref{eq:hills_eccentricity} predicts eccentricities marginally higher but broadly consistent with the fitted values. \citet{Cufari22} show that numerical methods indicate the peak of the probability density function of $e$ to be roughly consistent, although slightly higher, than the predictions of Equation \ref{eq:hills_eccentricity} and therefore our inferred orbits are reasonable assuming a suitable encounter between a WD binary and an IMBH occurs. These orbital properties will result in a fast runaway from initial binding to a final disruption implying the first stripping event occurred only shortly before disruption. As gravitational wave emission will change the orbit only very slowly, the initial orbit must also be similar to this final orbit. Together, these imply the WD was bound to the IMBH within a few tens of orbits of the final disruption.

It is difficult to quantify how likely such the encounter between a WD binary and an IMBH will be due to the poorly constrained rates of both. If we assume a limiting flux for \textit{Fermi}-GBM of $5\times10^{-8}$ erg cm$^{-2}$ s$^{-1}$ \citep[][see also \citet{Perley16}]{Liu25}, then triggers as luminous as GRB 250702B's peak should be detectable out to a luminosity distance of $\sim53.1$ Gpc. Accounting for time dilation effects, with only this one source uniquely observed over the 17 years of \textit{Fermi}'s mission, this suggests a volumetric rate of $\sim10^{-4}$ Gpc$^{-3}$ yr$^{-1}$, consistent with the estimates of \citet{Gompertz25}. From the other perspective, the assumed rate of typical TDEs \citep[$10^{-5}-10^{-4}$ galaxy$^{-1}$ yr$^{-1}$, see][and references therein]{Gezari21} represents an upper limit on events like this given the relative rates of observed IMBH TDEs\footnote{At time of writing, $\sim200$ to 300 TDE candidates are known of which two to three have been linked to IMBHs.}. The disruption of a WD is likely even rarer and there are no confirmed candidates observationally. However, \citet{Ye23} predict the capture rate of WDs by nuclear IMBHs to be of order $\sim10^{-6}$ galaxy$^{-1}$ yr$^{-1}$, consistent with estimates from \citet[][see also \citet{Maguire20} and references therein]{Macleod14}. For IMBHs in the outskirts of a galaxy, this is likely to be several orders of magnitude smaller especially given the eccentricity and beaming requirements we derive above. This implies a volumetric rate of order $<10^{-3}$ Gpc$^{-3}$ yr$^{-1}$ and consistent with the rarity implied by our calculated volumetric rate.

\section{Conclusions}
\label{sec:conclusions}

We have presented and analysed the results of X-ray observations of GRB 250702B and have also identified several key features from across the electromagnetic spectrum that must be explained in any model of this event. We have shown that all of these features can be exhibited by the tidal disruption of white dwarf by an intermediate mass black hole. These features and how they can arise from a WD-IMBH TDE are summarised below:
\begin{enumerate}
    \item \textit{Precursor X-ray emission from $\sim$few 10 ks prior to the gamma-ray flares.} The WD undergoes multiple pericentric passages due to its bound orbit. Material is repeatedly stripped prior to its final disruption and accreted to produce a rising X-ray counterpart.
    \item \textit{High peak X-ray luminosity that quickly fades as $\sim t^{-1.9}$.} Super-Eddington accretion rates result in the launch of a wide jet with a modest Lorentz factor. Internal processes produce the X-ray emission and the steeper slope than classically expected from a TDE is a result of the WD's eccentric orbit.
    \item \textit{A fast decaying NIR counterpart and long lived radio counterpart.} NIR emission is likely to be thermal emission from the accretion flow or some other accretion source which will fade quickly following the X-ray. The radio emission is the result of the jet shocking external material to produce a synchrotron afterglow.
    \item \textit{Gamma-ray flares peaking at hundreds of keV and stretching to MeV energies with fast ($\sim$seconds timescale) variability.} Stream-stream shocks occurring at a radius of $\sim$a lightsecond can produce variable flashes of X-ray photons which are upscattered by the jet. The modest $\Gamma$ of the jet is sufficient to achieve the MeV photon energies observed.
    \item \textit{Near constant intervals between the gamma-ray flare times.} The shocks occur as the stream or disk encounter each other at the time of the WD's pericentric passages. Such interactions do not occur on each pass due to the precession of one or more of these components.
\end{enumerate}

A thorough rate calculation is beyond the scope of this Letter, however, the expected rate is broadly compatible with that implied by the uniqueness of this event. While there is no conclusive proof that GRB 250702B was indeed a WD-IMBH TDE and several models remain eminently feasible, we have shown that it is possible within the data currently available to us. Future observations and analysis will help determine the exact properties of the jet and allow us to fully evaluate the inferences we have made here.

\section*{Acknowledgements}

We thank the referee for their insightful and thorough comments that have significantly improved this manuscript.

We thank Pat Slane and the Chandra X-ray Center team for approving and observing our DDT request for Chandra ObsID 31468. We also thank Brendan O'Connor for his note on ObsID 31011.

This work made use of data supplied by the UK Swift Science Data Centre at the University of Leicester.

BPG acknowledges support from STFC grant No. ST/Y002253/1 and The Leverhulme Trust grant No. RPG-2024-117.
PGJ is funded by the European Union (ERC, Starstruck, 101095973). Views and opinions expressed are however those of the authors only and do not necessarily reflect those of the European Union or the European Research Council Executive Agency. Neither the European Union nor the granting authority can be held responsible for them.
PTO acknowledges support from UKRI under grant ST/W000857/1.
JCR was supported by NASA through the NASA Hubble Fellowship grant \#HST-HF2-51587.001-A awarded by the Space Telescope Science Institute, which is operated by the Association of Universities for Research in Astronomy, Inc., for NASA, under contract NAS5-26555.
RLCS acknowledges support from The Leverhulme Trust grant RPG-2023-240.

For the purpose of open access, the author has applied a Creative Commons Attribution (CC BY) licence to the Author Accepted Manuscript version arising from this submission.

%%%%%%%%%%%%%%%%%%%%%%%%%%%%%%%%%%%%%%%%%%%%%%%%%%
\section*{Data Availability}

The data used in this work are publicly available, either from the UKSSDC or the Chandra Data Archive (see footnotes above for links), or from \citet{Li25}.



\bibliographystyle{mnras}
\bibliography{main}

@ARTICLE{AmaroSeoane18,
       author = {{Amaro-Seoane}, P.},
        title = "{Relativistic dynamics and extreme mass ratio inspirals}",
      journal = {Living Reviews in Relativity},
         year = 2018,
        month = dec,
       volume = {21},
       number = {1},
          eid = {4},
        pages = {4},
          doi = {10.1007/s41114-018-0013-8},
archivePrefix = {arXiv},
       eprint = {1205.5240},
 primaryClass = {astro-ph.CO},
       adsurl = {https://ui.adsabs.harvard.edu/abs/2018LRR....21....4A}
}

@ARTICLE{Ambrosino20,
       author = {{Ambrosino}, Federico},
        title = "{White Dwarf mass-radius relation}",
      journal = {arXiv e-prints},
         year = 2020,
        month = dec,
          eid = {arXiv:2012.01242},
        pages = {arXiv:2012.01242},
          doi = {10.48550/arXiv.2012.01242},
archivePrefix = {arXiv},
       eprint = {2012.01242},
 primaryClass = {astro-ph.HE},
       adsurl = {https://ui.adsabs.harvard.edu/abs/2020arXiv201201242A}
}

@ARTICLE{An25,
       author = {{An}, Tao},
        title = "{A precessing magnetic jet as the engine of GRB 250702B}",
      journal = {arXiv e-prints},
         year = 2025,
        month = nov,
          eid = {arXiv:2511.09850},
        pages = {arXiv:2511.09850},
          doi = {10.48550/arXiv.2511.09850},
archivePrefix = {arXiv},
       eprint = {2511.09850},
 primaryClass = {astro-ph.HE},
       adsurl = {https://ui.adsabs.harvard.edu/abs/2025arXiv251109850A}
}

@ARTICLE{Andalman22,
       author = {{Andalman}, Zachary L. and {Liska}, Matthew T.~P. and {Tchekhovskoy}, Alexander and {Coughlin}, Eric R. and {Stone}, Nicholas},
        title = "{Tidal disruption discs formed and fed by stream-stream and stream-disc interactions in global GRHD simulations}",
      journal = {\mnras},
         year = 2022,
        month = feb,
       volume = {510},
       number = {2},
        pages = {1627-1648},
          doi = {10.1093/mnras/stab3444},
archivePrefix = {arXiv},
       eprint = {2008.04922},
 primaryClass = {astro-ph.HE},
       adsurl = {https://ui.adsabs.harvard.edu/abs/2022MNRAS.510.1627A}
}

@INPROCEEDINGS{Arnaud96,
       author = {{Arnaud}, K.~A.},
        title = "{XSPEC: The First Ten Years}",
    booktitle = {Astronomical Data Analysis Software and Systems V},
         year = 1996,
       editor = {{Jacoby}, George H. and {Barnes}, Jeannette},
       series = {Astronomical Society of the Pacific Conference Series},
       volume = {101},
        month = jan,
        pages = {17},
       adsurl = {https://ui.adsabs.harvard.edu/abs/1996ASPC..101...17A}
}

@ARTICLE{Beniamini23,
       author = {{Beniamini}, Paz and {Piran}, Tsvi and {Matsumoto}, Tatsuya},
        title = "{Swift J1644+57 as an off-axis Jet}",
      journal = {\mnras},
         year = 2023,
        month = sep,
       volume = {524},
       number = {1},
        pages = {1386-1395},
          doi = {10.1093/mnras/stad1950},
archivePrefix = {arXiv},
       eprint = {2305.06370},
 primaryClass = {astro-ph.HE},
       adsurl = {https://ui.adsabs.harvard.edu/abs/2023MNRAS.524.1386B}
}

@ARTICLE{Beniamini25,
       author = {{Beniamini}, Paz and {Perets}, Hagai B. and {Granot}, Jonathan},
        title = "{Ultra-long Gamma-ray Bursts from Micro-Tidal Disruption Events: The Case of GRB 250702B}",
      journal = {arXiv e-prints},
         year = 2025,
        month = sep,
          eid = {arXiv:2509.22779},
        pages = {arXiv:2509.22779},
          doi = {10.48550/arXiv.2509.22779},
archivePrefix = {arXiv},
       eprint = {2509.22779},
 primaryClass = {astro-ph.HE},
       adsurl = {https://ui.adsabs.harvard.edu/abs/2025arXiv250922779B}
}

@ARTICLE{Bonnerot21b,
       author = {{Bonnerot}, C. and {Stone}, N.~C.},
        title = "{Formation of an Accretion Flow}",
      journal = {\ssr},
         year = 2021,
        month = feb,
       volume = {217},
       number = {1},
          eid = {16},
        pages = {16},
          doi = {10.1007/s11214-020-00789-1},
archivePrefix = {arXiv},
       eprint = {2008.11731},
 primaryClass = {astro-ph.HE},
       adsurl = {https://ui.adsabs.harvard.edu/abs/2021SSRv..217...16B}
}

@ARTICLE{Bonnerot21,
       author = {{Bonnerot}, Cl{\'e}ment and {Lu}, Wenbin and {Hopkins}, Philip F.},
        title = "{First light from tidal disruption events}",
      journal = {\mnras},
         year = 2021,
        month = jul,
       volume = {504},
       number = {4},
        pages = {4885-4905},
          doi = {10.1093/mnras/stab398},
archivePrefix = {arXiv},
       eprint = {2012.12271},
 primaryClass = {astro-ph.HE},
       adsurl = {https://ui.adsabs.harvard.edu/abs/2021MNRAS.504.4885B}
}

@ARTICLE{Carney25,
       author = {{Carney}, Jonathan and {Andreoni}, Igor and {O'Connor}, Brendan and {Freeburn}, James and {Skobe}, Hannah and {Westcott}, Lewi and {Busmann}, Malte and {Palmese}, Antonella and {Hall}, Xander J. and {Gill}, Ramandeep and {Beniamini}, Paz and {Coughlin}, Eric R. and {Kilpatrick}, Charles D. and {Anumarlapudi}, Akash and {Law}, Nicholas M. and {Corbett}, Hank and {Ahumada}, Tomas and {Chen}, Ping and {Conselice}, Christopher and {Damke}, Guillermo and {Das}, Kaustav K. and {Gal-Yam}, Avishay and {Gruen}, Daniel and {Heathcote}, Steve and {Hu}, Lei and {Karambelkar}, Viraj and {Kasliwal}, Mansi and {Labrie}, Kathleen and {Pasham}, Dheeraj and {Riffeser}, Arno and {Schmidt}, Michael and {Sharma}, Kritti and {Wilke}, Silona and {Zang}, Weicheng},
        title = "{Optical/Infrared Observations of the Extraordinary GRB 250702B: A Highly Obscured Afterglow in a Massive Galaxy Consistent with Multiple Possible Progenitors}",
      journal = {\apjl},
         year = 2025,
        month = dec,
       volume = {994},
       number = {2},
          eid = {L46},
        pages = {L46},
          doi = {10.3847/2041-8213/ae1d67},
archivePrefix = {arXiv},
       eprint = {2509.22784},
 primaryClass = {astro-ph.HE},
       adsurl = {https://ui.adsabs.harvard.edu/abs/2025ApJ...994L..46C}
}

@ARTICLE{Chen22,
       author = {{Chen}, Xian and {Qiu}, Yu and {Li}, Shuo and {Liu}, F.~K.},
        title = "{Milli-Hertz Gravitational-wave Background Produced by Quasiperiodic Eruptions}",
      journal = {\apj},
         year = 2022,
        month = may,
       volume = {930},
       number = {2},
          eid = {122},
        pages = {122},
          doi = {10.3847/1538-4357/ac63bf},
archivePrefix = {arXiv},
       eprint = {2112.03408},
 primaryClass = {astro-ph.HE},
       adsurl = {https://ui.adsabs.harvard.edu/abs/2022ApJ...930..122C}
}

@ARTICLE{Coughlin14,
       author = {{Coughlin}, Eric R. and {Begelman}, Mitchell C.},
        title = "{Hyperaccretion during Tidal Disruption Events: Weakly Bound Debris Envelopes and Jets}",
      journal = {\apj},
         year = 2014,
        month = feb,
       volume = {781},
       number = {2},
          eid = {82},
        pages = {82},
          doi = {10.1088/0004-637X/781/2/82},
archivePrefix = {arXiv},
       eprint = {1312.5314},
 primaryClass = {astro-ph.HE},
       adsurl = {https://ui.adsabs.harvard.edu/abs/2014ApJ...781...82C}
}

@ARTICLE{Crumley16,
       author = {{Crumley}, P. and {Lu}, W. and {Santana}, R. and {Hern{\'a}ndez}, R.~A. and {Kumar}, P. and {Markoff}, S.},
        title = "{Swift J1644+57: an ideal test bed of radiation mechanisms in a relativistic super-Eddington jet}",
      journal = {\mnras},
         year = 2016,
        month = jul,
       volume = {460},
       number = {1},
        pages = {396-416},
          doi = {10.1093/mnras/stw967},
archivePrefix = {arXiv},
       eprint = {1604.06468},
 primaryClass = {astro-ph.HE},
       adsurl = {https://ui.adsabs.harvard.edu/abs/2016MNRAS.460..396C}
}

@ARTICLE{Cufari22a,
       author = {{Cufari}, M. and {Coughlin}, Eric R. and {Nixon}, C.~J.},
        title = "{The Eccentric Nature of Eccentric Tidal Disruption Events}",
      journal = {\apj},
         year = 2022,
        month = jan,
       volume = {924},
       number = {1},
          eid = {34},
        pages = {34},
          doi = {10.3847/1538-4357/ac32be},
archivePrefix = {arXiv},
       eprint = {2110.11374},
 primaryClass = {astro-ph.HE},
       adsurl = {https://ui.adsabs.harvard.edu/abs/2022ApJ...924...34C}
}

@ARTICLE{Cufari22,
       author = {{Cufari}, M. and {Coughlin}, Eric R. and {Nixon}, C.~J.},
        title = "{Using the Hills Mechanism to Generate Repeating Partial Tidal Disruption Events and ASASSN-14ko}",
      journal = {\apjl},
         year = 2022,
        month = apr,
       volume = {929},
       number = {2},
          eid = {L20},
        pages = {L20},
          doi = {10.3847/2041-8213/ac6021},
archivePrefix = {arXiv},
       eprint = {2203.08162},
 primaryClass = {astro-ph.HE},
       adsurl = {https://ui.adsabs.harvard.edu/abs/2022ApJ...929L..20C}
}

@ARTICLE{Dai18,
       author = {{Dai}, Lixin and {McKinney}, Jonathan C. and {Roth}, Nathaniel and {Ramirez-Ruiz}, Enrico and {Miller}, M. Coleman},
        title = "{A Unified Model for Tidal Disruption Events}",
      journal = {\apjl},
         year = 2018,
        month = jun,
       volume = {859},
       number = {2},
          eid = {L20},
        pages = {L20},
          doi = {10.3847/2041-8213/aab429},
archivePrefix = {arXiv},
       eprint = {1803.03265},
 primaryClass = {astro-ph.HE},
       adsurl = {https://ui.adsabs.harvard.edu/abs/2018ApJ...859L..20D}
}

@ARTICLE{Darbha19,
       author = {{Darbha}, Siva and {Coughlin}, Eric R. and {Kasen}, Daniel and {Nixon}, Chris},
        title = "{Ultra-deep tidal disruption events: prompt self-intersections and observables}",
      journal = {\mnras},
         year = 2019,
        month = oct,
       volume = {488},
       number = {4},
        pages = {5267-5278},
          doi = {10.1093/mnras/stz1923},
archivePrefix = {arXiv},
       eprint = {1905.07056},
 primaryClass = {astro-ph.HE},
       adsurl = {https://ui.adsabs.harvard.edu/abs/2019MNRAS.488.5267D}
}

@ARTICLE{Evans09,
       author = {{Evans}, P.~A. and {Beardmore}, A.~P. and {Page}, K.~L. and {Osborne}, J.~P. and {O'Brien}, P.~T. and {Willingale}, R. and {Starling}, R.~L.~C. and {Burrows}, D.~N. and {Godet}, O. and {Vetere}, L. and {Racusin}, J. and {Goad}, M.~R. and {Wiersema}, K. and {Angelini}, L. and {Capalbi}, M. and {Chincarini}, G. and {Gehrels}, N. and {Kennea}, J.~A. and {Margutti}, R. and {Morris}, D.~C. and {Mountford}, C.~J. and {Pagani}, C. and {Perri}, M. and {Romano}, P. and {Tanvir}, N.},
        title = "{Methods and results of an automatic analysis of a complete sample of Swift-XRT observations of GRBs}",
      journal = {\mnras},
         year = 2009,
        month = aug,
       volume = {397},
       number = {3},
        pages = {1177-1201},
          doi = {10.1111/j.1365-2966.2009.14913.x},
archivePrefix = {arXiv},
       eprint = {0812.3662},
 primaryClass = {astro-ph},
       adsurl = {https://ui.adsabs.harvard.edu/abs/2009MNRAS.397.1177E}
}

@ARTICLE{EylesFerris22,
       author = {{Eyles-Ferris}, R.~A.~J. and {Starling}, R.~L.~C. and {O'Brien}, P.~T. and {Nixon}, C.~J. and {Coughlin}, Eric R.},
        title = "{Simulated optical light curves of super-Eddington tidal disruption events with ZEBRA flows}",
      journal = {\mnras},
         year = 2022,
        month = dec,
       volume = {517},
       number = {4},
        pages = {6013-6021},
          doi = {10.1093/mnras/stac3073},
archivePrefix = {arXiv},
       eprint = {2210.12168},
 primaryClass = {astro-ph.HE},
       adsurl = {https://ui.adsabs.harvard.edu/abs/2022MNRAS.517.6013E}
}

@ARTICLE{EylesFerris24,
       author = {{Eyles-Ferris}, R.~A.~J. and {Nixon}, C.~J. and {Coughlin}, E.~R. and {O'Brien}, P.~T.},
        title = "{Ultradeep Cover: An Exotic and Jetted Tidal Disruption Event Candidate Disguised as a Gamma-Ray Burst}",
      journal = {\apjl},
         year = 2024,
        month = apr,
       volume = {965},
       number = {2},
          eid = {L20},
        pages = {L20},
          doi = {10.3847/2041-8213/ad3922},
archivePrefix = {arXiv},
       eprint = {2404.03982},
 primaryClass = {astro-ph.HE},
       adsurl = {https://ui.adsabs.harvard.edu/abs/2024ApJ...965L..20E}
}

@ARTICLE{ForemanMackey13,
       author = {{Foreman-Mackey}, Daniel and {Hogg}, David W. and {Lang}, Dustin and {Goodman}, Jonathan},
        title = "{emcee: The MCMC Hammer}",
      journal = {\pasp},
         year = 2013,
        month = mar,
       volume = {125},
       number = {925},
        pages = {306},
          doi = {10.1086/670067},
archivePrefix = {arXiv},
       eprint = {1202.3665},
 primaryClass = {astro-ph.IM},
       adsurl = {https://ui.adsabs.harvard.edu/abs/2013PASP..125..306F}
}

@ARTICLE{Frail01,
       author = {{Frail}, D.~A. and {Kulkarni}, S.~R. and {Sari}, R. and {Djorgovski}, S.~G. and {Bloom}, J.~S. and {Galama}, T.~J. and {Reichart}, D.~E. and {Berger}, E. and {Harrison}, F.~A. and {Price}, P.~A. and {Yost}, S.~A. and {Diercks}, A. and {Goodrich}, R.~W. and {Chaffee}, F.},
        title = "{Beaming in Gamma-Ray Bursts: Evidence for a Standard Energy Reservoir}",
      journal = {\apjl},
         year = 2001,
        month = nov,
       volume = {562},
       number = {1},
        pages = {L55-L58},
          doi = {10.1086/338119},
archivePrefix = {arXiv},
       eprint = {astro-ph/0102282},
 primaryClass = {astro-ph},
       adsurl = {https://ui.adsabs.harvard.edu/abs/2001ApJ...562L..55F}
}

@ARTICLE{Franchini16,
       author = {{Franchini}, Alessia and {Lodato}, Giuseppe and {Facchini}, Stefano},
        title = "{Lense-Thirring precession around supermassive black holes during tidal disruption events}",
      journal = {\mnras},
         year = 2016,
        month = jan,
       volume = {455},
       number = {2},
        pages = {1946-1956},
          doi = {10.1093/mnras/stv2417},
archivePrefix = {arXiv},
       eprint = {1510.04879},
 primaryClass = {astro-ph.HE},
       adsurl = {https://ui.adsabs.harvard.edu/abs/2016MNRAS.455.1946F}
}

@INPROCEEDINGS{Fruscione06,
       author = {{Fruscione}, Antonella and {McDowell}, Jonathan C. and {Allen}, Glenn E. and {Brickhouse}, Nancy S. and {Burke}, Douglas J. and {Davis}, John E. and {Durham}, Nick and {Elvis}, Martin and {Galle}, Elizabeth C. and {Harris}, Daniel E. and {Huenemoerder}, David P. and {Houck}, John C. and {Ishibashi}, Bish and {Karovska}, Margarita and {Nicastro}, Fabrizio and {Noble}, Michael S. and {Nowak}, Michael A. and {Primini}, Frank A. and {Siemiginowska}, Aneta and {Smith}, Randall K. and {Wise}, Michael},
        title = "{CIAO: Chandra's data analysis system}",
    booktitle = {Observatory Operations: Strategies, Processes, and Systems},
         year = 2006,
       editor = {{Silva}, David R. and {Doxsey}, Rodger E.},
       series = {Society of Photo-Optical Instrumentation Engineers (SPIE) Conference Series},
       volume = {6270},
        month = jun,
          eid = {62701V},
        pages = {62701V},
          doi = {10.1117/12.671760},
       adsurl = {https://ui.adsabs.harvard.edu/abs/2006SPIE.6270E..1VF}
}

@ARTICLE{GCN40883,
       author = {{Fermi GBM Team}},
        title = "{GRB 250702B: Fermi GBM Final Real-time Localization}",
      journal = {GRB Coordinates Network},
         year = 2025,
        month = jul,
       volume = {40883},
        pages = {1},
       adsurl = {https://ui.adsabs.harvard.edu/abs/2025GCN.40883....1F}
}

@ARTICLE{GCN40886,
       author = {{Fermi GBM Team}},
        title = "{GRB 250702D: Fermi GBM Final Localization Correction}",
      journal = {GRB Coordinates Network},
         year = 2025,
        month = jul,
       volume = {40886},
        pages = {1},
       adsurl = {https://ui.adsabs.harvard.edu/abs/2025GCN.40886....1F}
}

@ARTICLE{GCN40890,
       author = {{Fermi GBM Team}},
        title = "{GRB 250702E: Fermi GBM Final Real-time Localization}",
      journal = {GRB Coordinates Network},
         year = 2025,
        month = jul,
       volume = {40890},
        pages = {1},
       adsurl = {https://ui.adsabs.harvard.edu/abs/2025GCN.40890....1F}
}

@ARTICLE{GCN40891,
       author = {{Neights}, E. and {Roberts}, O.~J. and {Burns}, E. and {Veres}, P. and {Fermi-GBM Team}},
        title = "{Fermi GBM Triggers 250702B, C, D and E are likely from the same source}",
      journal = {GRB Coordinates Network},
         year = 2025,
        month = jul,
       volume = {40891},
        pages = {1},
       adsurl = {https://ui.adsabs.harvard.edu/abs/2025GCN.40891....1N}
}

@ARTICLE{GCN40903,
       author = {{DeLaunay}, James and {Ronchini}, Samuele and {Tohuvavohu}, Aaron and {Kennea}, Jamie A. and {Parsotan}, Tyler and {Williams}, Maia},
        title = "{GRB 250702D, C, E: Swift/BAT-GUANO localization skymap of a burst or galactic transient}",
      journal = {GRB Coordinates Network},
         year = 2025,
        month = jul,
       volume = {40903},
        pages = {1},
       adsurl = {https://ui.adsabs.harvard.edu/abs/2025GCN.40903....1D}
}

@ARTICLE{GCN40906,
       author = {{Cheng}, H.~Q. and {Zhao}, G.~Y. and {Zhou}, C. and {Cheng}, Y.~H. and {Zhang}, Y.~J. and {Hu}, J.~W. and {Sun}, H. and {Ling}, Z.~X. and {Einstein Probe Team}},
        title = "{EP250702a : an X-ray transient detected by Einstein Probe likely associated with GRB 250702B,C,D,E}",
      journal = {GRB Coordinates Network},
         year = 2025,
        month = jul,
       volume = {40906},
        pages = {1},
       adsurl = {https://ui.adsabs.harvard.edu/abs/2025GCN.40906....1C}
}

@ARTICLE{GCN40910,
       author = {{Kawakubo}, Y. and {Serino}, M. and {Negoro}, H. and {Nakajima}, M. and {Takagi}, K. and {Takahashi}, H. and {Tatano}, K. and {Nishio}, H. and {Mihara}, T. and {Tamagawa}, T. and {Kawai}, N. and {Matsuoka}, M. and {Sakamoto}, T. and {Sugita}, S. and {Hiramatsu}, H. and {Kondo}, Y. and {Yoshida}, A. and {Tsuboi}, Y. and {Sugai}, H. and {Nagashima}, N. and {Shidatsu}, M. and {Niida}, Y. and {Kang}, C. and {Nakamoto}, T. and {Takahashi}, I. and {Yatsu}, Y. and {Nakahira}, S. and {Ueno}, S. and {Tomida}, H. and {Ishikawa}, M. and {Ogawa}, S. and {Kurihara}, M. and {Ueda}, Y. and {Fujiwara}, K. and {Yamauchi}, M. and {Nishio}, M. and {Hiraizumi}, C. and {Yamaoka}, K. and {Sugizaki}, M. and {Iwakiri}, W. and {Kawamuro}, T. and {Yamada}, S.},
        title = "{MAXI/GSC detection of an X-ray activity from a transient associated with GRB 250702B,C,D,E and EP250702a}",
      journal = {GRB Coordinates Network},
         year = 2025,
        month = jul,
       volume = {40910},
        pages = {1},
       adsurl = {https://ui.adsabs.harvard.edu/abs/2025GCN.40910....1K}
}

@ARTICLE{GCN40914,
       author = {{Frederiks}, D. and {Lysenko}, A. and {Ridnaia}, A. and {Svinkin}, D. and {Tsvetkova}, A. and {Ulanov}, M. and {Cline}, T. and {Konus-Wind Team}},
        title = "{GRBs 250702B,C,D,E / EP250702a: Konus-Wind detection of a hard X-ray transient activity}",
      journal = {GRB Coordinates Network},
         year = 2025,
        month = jul,
       volume = {40914},
        pages = {1},
       adsurl = {https://ui.adsabs.harvard.edu/abs/2025GCN.40914....1F}
}

@ARTICLE{GCN40917,
       author = {{Cheng}, H.~Q. and {Zhang}, Y.~J. and {Zhao}, G.~Y. and {Zhou}, C. and {Cheng}, Y.~H. and {Wang}, Y.~L. and {Coti Zelati}, F. and {Marino}, A. and {Rea}, N. and {Hu}, J.~W. and {Sun}, H. and {Ling}, Z.~X. and {Yuan}, W. and {Einstein Probe Team}},
        title = "{EP250702a/GRB 250702B, C, D, E: EP-FXT follow-up observation}",
      journal = {GRB Coordinates Network},
         year = 2025,
        month = jul,
       volume = {40917},
        pages = {1},
       adsurl = {https://ui.adsabs.harvard.edu/abs/2025GCN.40917....1C}
}

@ARTICLE{GCN40919,
       author = {{Kennea}, J.~A. and {Siegel}, M.~H. and {Evans}, P.~A. and {Page}, K.~L. and {O'Connor}, B. and {Swift Team}},
        title = "{GRBs 250702B/C/D/E / EP250702a: Swift XRT localization}",
      journal = {GRB Coordinates Network},
         year = 2025,
        month = jul,
       volume = {40919},
        pages = {1},
       adsurl = {https://ui.adsabs.harvard.edu/abs/2025GCN.40919....1K}
}

@ARTICLE{GCN40923,
       author = {{SVOM/GRM Team} and {Wang}, Chen-Wei and {Zheng}, Shi-Jie and {Huang}, Yue and {Xiong}, Shao-Lin and {Zhang}, Shuang-Nan and {Svom/Eclairs Team} and {Coleiro}, Alexis and {Svom Team}},
        title = "{GRBs 250702B,C,D,E / EP250702a: SVOM/GRM observation}",
      journal = {GRB Coordinates Network},
         year = 2025,
        month = jul,
       volume = {40923},
        pages = {1},
       adsurl = {https://ui.adsabs.harvard.edu/abs/2025GCN.40923....1S}
}

@ARTICLE{GCN40924,
       author = {{Martin-Carrillo}, A. and {Levan}, A.~J. and {Schneider}, B. and {An}, J. and {Xu}, D. and {Campana}, S. and {Corcoran}, G. and {Jonker}, P.~G. and {Malesani}, D.~B. and {Stargate Collaboration}},
        title = "{GRBs 250702B,C,D,E / EP250702a: VLT/HAWK-I NIR candidate}",
      journal = {GRB Coordinates Network},
         year = 2025,
        month = jul,
       volume = {40924},
        pages = {1},
       adsurl = {https://ui.adsabs.harvard.edu/abs/2025GCN.40924....1M}
}

@ARTICLE{GCN40961,
       author = {{Levan}, A.~J. and {Martin-Carrillo}, A. and {Schneider}, B. and {Jonker}, P.~G. and {Malesani}, D.~B. and {Gompertz}, B.~P. and {Corcoran}, G. and {De Pasquale}, M. and {Stargate Collaboration}},
        title = "{GRBs 250702B,D,E / EP250702a: fast fading, extremely red counterpart}",
      journal = {GRB Coordinates Network},
         year = 2025,
        month = jul,
       volume = {40961},
        pages = {1},
       adsurl = {https://ui.adsabs.harvard.edu/abs/2025GCN.40961....1L}
}

@ARTICLE{GCN40979,
       author = {{Sfaradi}, Authors I. and {Margutti}, R. and {Farah}, W. and {Sheikh}, S. and {Chornock}, R. and {Garcia Lopez}, V. and {Bright}, J. and {Alexander}, K. and {Siemion}, A. and {Pollak}, A. and {Yao}, Y. and {Nayana A.}, J. and {Sears}, H.},
        title = "{Radio observation of GRBs 250702B,C,E / EP250702a with the Allen Telescope Array}",
      journal = {GRB Coordinates Network},
         year = 2025,
        month = jul,
       volume = {40979},
        pages = {1},
       adsurl = {https://ui.adsabs.harvard.edu/abs/2025GCN.40979....1S}
}

@ARTICLE{GCN40985,
       author = {{Bright}, Authors J. and {Carotenuto}, F. and {Jonker}, P.~G.},
        title = "{EP250702a/GRB250702 B,E,D: MeerKAT radio counterpart}",
      journal = {GRB Coordinates Network},
         year = 2025,
        month = jul,
       volume = {40985},
        pages = {1},
       adsurl = {https://ui.adsabs.harvard.edu/abs/2025GCN.40985....1B}
}

@ARTICLE{GCN41014,
       author = {{O'Connor}, Brendan and {Pasham}, Dheeraj and {Andreoni}, Igor and {Hare}, Jeremy},
        title = "{EP250702a/ GRB 250702B: NuSTAR X-ray Observations}",
      journal = {GRB Coordinates Network},
         year = 2025,
        month = jul,
       volume = {41014},
        pages = {1},
       adsurl = {https://ui.adsabs.harvard.edu/abs/2025GCN.41014....1O}
}

@ARTICLE{GCN41053,
       author = {{Sfaradi}, Authors I. and {Yao}, Y. and {Sears}, H. and {Wiston}, E. and {Margutti}, R. and {Chornock}, R. and {Alexander}, K.~D. and {Laskar}, T. and {Hammerstein}, E. and {Lu}, W.},
        title = "{GRB 250702B,D,E / EP250702a: 10 GHz detection with the VLA}",
      journal = {GRB Coordinates Network},
         year = 2025,
        month = jul,
       volume = {41053},
        pages = {1},
       adsurl = {https://ui.adsabs.harvard.edu/abs/2025GCN.41053....1S}
}

@ARTICLE{GCN41054,
       author = {{Atri}, Pikky and {Rhodes}, Lauren and {Fender}, Rob and {Hughes}, Andrew and {Motta}, Sara and {XKAT Collaboration}},
        title = "{GRB 250702B,D,E/EP250702A: MeerKAT radio observations at 1.28GHz}",
      journal = {GRB Coordinates Network},
         year = 2025,
        month = jul,
       volume = {41054},
        pages = {1},
       adsurl = {https://ui.adsabs.harvard.edu/abs/2025GCN.41054....1A}
}

@ARTICLE{GCN41059,
       author = {{Alexander}, Kate D. and {Miller-Jones}, James and {Goodwin}, Adelle and {Franz}, Noah and {Margutti}, Raffaella and {Chornock}, Ryan and {Pasham}, Dheeraj and {Berger}, Edo and {Cendes}, Yvette and {Christy}, Collin},
        title = "{GRB 250702B,D,E / EP250702a: ALMA detection}",
      journal = {GRB Coordinates Network},
         year = 2025,
        month = jul,
       volume = {41059},
        pages = {1},
       adsurl = {https://ui.adsabs.harvard.edu/abs/2025GCN.41059....1A}
}

@ARTICLE{GCN41061,
       author = {{Tetarenko}, A.~J. and {Bright}, J. and {Bower}, G. and {Graves}, S.},
        title = "{GRB 250702B,D,E/EP250702A: JCMT sub-mm observations at 350 GHz}",
      journal = {GRB Coordinates Network},
         year = 2025,
        month = jul,
       volume = {41061},
        pages = {1},
       adsurl = {https://ui.adsabs.harvard.edu/abs/2025GCN.41061....1T}
}

@ARTICLE{GCN41096,
       author = {{Levan}, A.~J. and {Martin-Carrillo}, A. and {Jonker}, P.~G. and {Malesani}, D.~B. and {Saccardi}, A. and {O'Brien}, P. and {Eyles-Ferris}, A.~J. and {de Ugarte Postigo}, A. and {Corcoran}, G. and {Vergani}, S.~D. and {Tanvir}, N.~R. and {Gompertz}, B.~P. and {Stargate collaboration}},
        title = "{EP250702a / GRB 250702B,D,E: Hubble Space Telescope Observations}",
      journal = {GRB Coordinates Network},
         year = 2025,
        month = jul,
       volume = {41096},
        pages = {1},
       adsurl = {https://ui.adsabs.harvard.edu/abs/2025GCN.41096....1G}
}

@ARTICLE{GCN41145,
       author = {{Balasubramanian}, A. and {Resmi}, L. and {Eappachen}, D. and {Jagan}, S.~K. and {Bhalerao}, V. and {Zhang}, B. and {Anupama}, G.~C. and {Sun}, H. and {Sahu}, D.~K. and {Yuan}, W.},
        title = "{EP250702a/GRB250702 B,D,E: uGMRT Radio detection in 1.26 GHz}",
      journal = {GRB Coordinates Network},
         year = 2025,
        month = jul,
       volume = {41145},
        pages = {1},
       adsurl = {https://ui.adsabs.harvard.edu/abs/2025GCN.41145....1B}
}

@ARTICLE{GCN41147,
       author = {{Grollimund}, N. and {Corbel}, S. and {Coleiro}, A. and {Cangemi}, F. and {Rodriguez}, J.},
        title = "{EP250702a/GRB250702 B,E,D: MeerKAT radio observations at 3.06 GHz}",
      journal = {GRB Coordinates Network},
         year = 2025,
        month = jul,
       volume = {41147},
        pages = {1},
       adsurl = {https://ui.adsabs.harvard.edu/abs/2025GCN.41147....1G}
}

@ARTICLE{GCN41215,
       author = {{Rhodes}, L. and {Atri}, P. and {Bright}, J.~S. and {Carotenuto}, F. and {Gurwell}, M. and {Keating}, G.~K. and {Sarin}, N.},
        title = "{GRB 250702B: SMA observations}",
      journal = {GRB Coordinates Network},
         year = 2025,
        month = aug,
       volume = {41215},
        pages = {1},
       adsurl = {https://ui.adsabs.harvard.edu/abs/2025GCN.41215....1R}
}

@ARTICLE{GCN41309,
       author = {{O'Connor}, B. and {Pasham}, D. and {Andreoni}, I. and {Hare}, J. and {Hall}, X. and {Carney}, J. and {Palmese}, A. and {Busmann}, M. and {Gruen}, D.},
        title = "{GRB 250702B: Chandra X-ray Detection}",
      journal = {GRB Coordinates Network},
         year = 2025,
        month = aug,
       volume = {41309},
        pages = {1},
       adsurl = {https://ui.adsabs.harvard.edu/abs/2025GCN.41309....1O}
}

@ARTICLE{GCN41767,
       author = {{Eyles-Ferris}, R. A. J. and {Levan}, A. J. and {Martin-Carrillo}, A. and {O'Brien}, P. T. and {De Pasquale}, M. and {Gompertz}, B. P. and {Laskar}, T. and {Malesani}, D. B. and {Rastinejad}, J. C. and {Schulze}, S. and {Tanvir}, N. R. and {Jonker}, P. G. and {Watson}, D.},
        title = "{GRB 250702B: Late time Chandra observations}",
      journal = {GRB Coordinates Network},
         year = 2025,
        month = sep,
       volume = {413767},
        pages = {1},
       adsurl = {https://ui.adsabs.harvard.edu/abs/2025GCN.41767....1E}
}

@ARTICLE{Gezari21,
       author = {{Gezari}, Suvi},
        title = "{Tidal Disruption Events}",
      journal = {\araa},
         year = 2021,
        month = sep,
       volume = {59},
        pages = {21-58},
          doi = {10.1146/annurev-astro-111720-030029},
archivePrefix = {arXiv},
       eprint = {2104.14580},
 primaryClass = {astro-ph.HE},
       adsurl = {https://ui.adsabs.harvard.edu/abs/2021ARA&A..59...21G}
}

@ARTICLE{Gompertz25,
       author = {{Gompertz}, Benjamin P. and {Levan}, Andrew J. and {Laskar}, Tanmoy and {Schneider}, Benjamin and {Chrimes}, Ashley A. and {Martin-Carrillo}, Antonio and {Sneppen}, Albert and {ONeill}, David and {Malesani}, Daniele B. and {Jonker}, Peter G. and {Burns}, Eric and {Corcoran}, Gregory and {Cotter}, Laura and {de Ugarte Postigo}, Antonio and {Dimple} and {Eyles-Ferris}, Rob A.~J. and {Izzo}, L. and {Jakobsson}, Pall and {Lamb}, Gavin P. and {Palmerio}, Jesse T. and {Pugliese}, Giovanna and {Edvige Ravasio}, Maria and {Saccardi}, Andrea and {Salvaterra}, Ruben and {Sarin}, Nikhil and {Schulze}, Steve and {Tanvir}, Nial and {Wortley}, Makenzie E.},
        title = "{JWST Spectroscopy of GRB 250702B: An Extremely Rare and Exceptionally Energetic Burst in a Dusty, Massive Galaxy at $z=1.036$}",
      journal = {arXiv e-prints},
         year = 2025,
        month = sep,
          eid = {arXiv:2509.22778},
        pages = {arXiv:2509.22778},
          doi = {10.48550/arXiv.2509.22778},
archivePrefix = {arXiv},
       eprint = {2509.22778},
 primaryClass = {astro-ph.HE},
       adsurl = {https://ui.adsabs.harvard.edu/abs/2025arXiv250922778G}
}

@ARTICLE{Hammerstein25,
       author = {{Hammerstein}, Erica and {Cenko}, S. Bradley and {Andreoni}, Igor and {Charalampopoulos}, Panos and {Chornock}, Ryan and {Margutti}, Raffaella and {O'Connor}, Brendan and {Schulze}, Steve and {Sollerman}, Jesper and {Barway}, Sudhanshu and {Bhalerao}, Varun and {Anupama G.}, C. and {Kumar}, Harsh and {Paris}, Ester Marini. Diego and {Perley}, Daniel A. and {Rossi}, Andrea and {Yao}, Yuhan},
        title = "{The Jetted Tidal Disruption Event AT2022cmc: Investigating Connections to the Optical Tidal Disruption Event Population and Spectral Subclasses Through Late-Time Follow-up}",
      journal = {arXiv e-prints},
         year = 2025,
        month = jun,
          eid = {arXiv:2506.08250},
        pages = {arXiv:2506.08250},
          doi = {10.48550/arXiv.2506.08250},
archivePrefix = {arXiv},
       eprint = {2506.08250},
 primaryClass = {astro-ph.HE},
       adsurl = {https://ui.adsabs.harvard.edu/abs/2025arXiv250608250H}
}

@ARTICLE{Hayasaki18,
       author = {{Hayasaki}, Kimitake and {Zhong}, Shiyan and {Li}, Shuo and {Berczik}, Peter and {Spurzem}, Rainer},
        title = "{Classification of Tidal Disruption Events Based on Stellar Orbital Properties}",
      journal = {\apj},
         year = 2018,
        month = mar,
       volume = {855},
       number = {2},
          eid = {129},
        pages = {129},
          doi = {10.3847/1538-4357/aab0a5},
archivePrefix = {arXiv},
       eprint = {1802.06798},
 primaryClass = {astro-ph.HE},
       adsurl = {https://ui.adsabs.harvard.edu/abs/2018ApJ...855..129H}
}

@ARTICLE{Hills88,
       author = {{Hills}, J.~G.},
        title = "{Hyper-velocity and tidal stars from binaries disrupted by a massive Galactic black hole}",
      journal = {\nat},
         year = 1988,
        month = feb,
       volume = {331},
       number = {6158},
        pages = {687-689},
          doi = {10.1038/331687a0},
       adsurl = {https://ui.adsabs.harvard.edu/abs/1988Natur.331..687H}
}

@ARTICLE{Jonker13,
       author = {{Jonker}, P.~G. and {Glennie}, A. and {Heida}, M. and {Maccarone}, T. and {Hodgkin}, S. and {Nelemans}, G. and {Miller-Jones}, J.~C.~A. and {Torres}, M.~A.~P. and {Fender}, R.},
        title = "{Discovery of a New Kind of Explosive X-Ray Transient near M86}",
      journal = {\apj},
         year = 2013,
        month = dec,
       volume = {779},
       number = {1},
          eid = {14},
        pages = {14},
          doi = {10.1088/0004-637X/779/1/14},
archivePrefix = {arXiv},
       eprint = {1310.7238},
 primaryClass = {astro-ph.HE},
       adsurl = {https://ui.adsabs.harvard.edu/abs/2013ApJ...779...14J}
}

@ARTICLE{King20,
       author = {{King}, Andrew},
        title = "{GSN 069 - A tidal disruption near miss}",
      journal = {\mnras},
         year = 2020,
        month = mar,
       volume = {493},
       number = {1},
        pages = {L120-L123},
          doi = {10.1093/mnrasl/slaa020},
archivePrefix = {arXiv},
       eprint = {2002.00970},
 primaryClass = {astro-ph.HE},
       adsurl = {https://ui.adsabs.harvard.edu/abs/2020MNRAS.493L.120K}
}

@ARTICLE{King22,
       author = {{King}, Andrew},
        title = "{Quasi-periodic eruptions from galaxy nuclei}",
      journal = {\mnras},
         year = 2022,
        month = sep,
       volume = {515},
       number = {3},
        pages = {4344-4349},
          doi = {10.1093/mnras/stac1641},
archivePrefix = {arXiv},
       eprint = {2206.04698},
 primaryClass = {astro-ph.GA},
       adsurl = {https://ui.adsabs.harvard.edu/abs/2022MNRAS.515.4344K}
}

@ARTICLE{Levan14,
       author = {{Levan}, A.~J. and {Tanvir}, N.~R. and {Starling}, R.~L.~C. and {Wiersema}, K. and {Page}, K.~L. and {Perley}, D.~A. and {Schulze}, S. and {Wynn}, G.~A. and {Chornock}, R. and {Hjorth}, J. and {Cenko}, S.~B. and {Fruchter}, A.~S. and {O'Brien}, P.~T. and {Brown}, G.~C. and {Tunnicliffe}, R.~L. and {Malesani}, D. and {Jakobsson}, P. and {Watson}, D. and {Berger}, E. and {Bersier}, D. and {Cobb}, B.~E. and {Covino}, S. and {Cucchiara}, A. and {de Ugarte Postigo}, A. and {Fox}, D.~B. and {Gal-Yam}, A. and {Goldoni}, P. and {Gorosabel}, J. and {Kaper}, L. and {Kr{\"u}hler}, T. and {Karjalainen}, R. and {Osborne}, J.~P. and {Pian}, E. and {S{\'a}nchez-Ram{\'\i}rez}, R. and {Schmidt}, B. and {Skillen}, I. and {Tagliaferri}, G. and {Th{\"o}ne}, C. and {Vaduvescu}, O. and {Wijers}, R.~A.~M.~J. and {Zauderer}, B.~A.},
        title = "{A New Population of Ultra-long Duration Gamma-Ray Bursts}",
      journal = {\apj},
         year = 2014,
        month = jan,
       volume = {781},
       number = {1},
          eid = {13},
        pages = {13},
          doi = {10.1088/0004-637X/781/1/13},
archivePrefix = {arXiv},
       eprint = {1302.2352},
 primaryClass = {astro-ph.HE},
       adsurl = {https://ui.adsabs.harvard.edu/abs/2014ApJ...781...13L}
}

@ARTICLE{Levan25,
       author = {{Levan}, Andrew J. and {Martin-Carrillo}, Antonio and {Laskar}, Tanmoy and {Eyles-Ferris}, Rob A.~J. and {Sneppen}, Albert and {Ravasio}, Maria Edvige and {Rastinejad}, Jillian C. and {Bright}, Joe S. and {Carotenuto}, Francesco and {Chrimes}, Ashley A. and {Corcoran}, Gregory and {Gompertz}, Benjamin P. and {Jonker}, Peter G. and {Lamb}, Gavin P. and {Malesani}, Daniele B. and {Saccardi}, Andrea and {S{\'a}nchez-Sierras}, Javier and {Schneider}, Benjamin and {Schulze}, Steve and {Tanvir}, Nial R. and {Vergani}, Susanna D. and {Watson}, Darach and {An}, Jie and {Bauer}, Franz E. and {Campana}, Sergio and {Cotter}, Laura and {van Dalen}, Joyce N.~D. and {D'Elia}, Valerio and {De Pasquale}, Massimiliano and {de Ugarte Postigo}, Antonio and {Dimple} and {Hartmann}, Dieter H. and {Hjorth}, Jens and {Izzo}, Luca and {Jakobsson}, P{\'a}ll and {Kumar}, Amit and {Melandri}, Andrea and {O'Brien}, Paul and {Piranomonte}, Silvia and {Pugliese}, Giovanna and {Quirola-V{\'a}squez}, Jonathan and {Starling}, Rhaana and {Tagliaferri}, Gianpiero and {Xu}, Dong and {Wortley}, Makenzie E.},
        title = "{The Day-long, Repeating GRB 250702B: A Unique Extragalactic Transient}",
      journal = {\apjl},
         year = 2025,
        month = sep,
       volume = {990},
       number = {1},
          eid = {L28},
        pages = {L28},
          doi = {10.3847/2041-8213/adf8e1},
       adsurl = {https://ui.adsabs.harvard.edu/abs/2025ApJ...990L..28L}
}

@ARTICLE{Li25,
       author = {{Li}, D.-Y. and {Zhang}, W.-D. and {Yang}, J. and {Chen}, J.-H. and {Yuan}, W. and {Cheng}, H.-Q. and {Xu}, F. and {Shu}, X.-W. and {Shen}, R.-F. and {Jiang}, N. and {Zhu}, J.-Z. and {Zhou}, C. and {Lei}, W.-H. and {Sun}, H. and {Jin}, C.-C. and {Dai}, L.-X. and {Zhang}, B. and {Yang}, Y.-H. and {Zhang}, W.-J. and {Feng}, H. and {Liu}, B.-F. and {Zhou}, H.-Y. and {Pan}, H.-W. and {Liu}, M.-J. and {Corbel}, S. and {Jagan}, S.~K. and {Baglio}, M.~C. and {Burns}, C. and {Cangemi}, F. and {Chen}, C. and {Cheng}, Y.-H. and {Coleiro}, A. and {Coti Zelati}, F. and {Das}, S.~R. and {Dong}, Z.-N. and {Galbany}, L. and {Grollimund}, N. and {Kelson}, D. and {Lai}, D. and {Li}, X. and {Liu}, Y. and {Marino}, A. and {Mockler}, B. and {O'Brien}, P. and {Qiao}, E.-L. and {Rea}, N. and {Resmi}, L. and {Rodriguez}, J. and {Saxton}, R. and {Sun}, L.-M. and {Tao}, L. and {Wang}, T.-G. and {Wang}, Y.-L. and {Wu}, X.-F. and {Xu}, D. and {Zhang}, Y.-J. and {Zhao}, G.-Y. and {Cai}, Z.-M. and {Chen}, Y. and {Cordier}, B. and {Fan}, Z. and {Gao}, H. and {Ghirlanda}, G. and {Hu}, J.-W. and {Huang}, Y.-F. and {Jia}, S.-M. and {Komossa}, S. and {Liu}, H.-Y. and {Liu}, H.-Q. and {Ness}, J.-U. and {Rau}, A. and {Sanders}, J. and {Soria}, R. and {Sun}, S.-L. and {Sun}, X.-J. and {Troja}, E. and {Wen}, S.-X. and {Xue}, Y.-Q. and {Yin}, Y.-H.~I. and {Zhang}, C. and {Zhang}, S.-N. and {Zhang}, Y.-H.},
        title = "{A fast powerful X-ray transient from possible tidal disruption of a white dwarf}",
      journal = {arXiv e-prints},
         year = 2025,
        month = sep,
          eid = {arXiv:2509.25877},
        pages = {arXiv:2509.25877},
          doi = {10.48550/arXiv.2509.25877},
archivePrefix = {arXiv},
       eprint = {2509.25877},
 primaryClass = {astro-ph.HE},
       adsurl = {https://ui.adsabs.harvard.edu/abs/2025arXiv250925877L}
}

@ARTICLE{Lien16,
       author = {{Lien}, Amy and {Sakamoto}, Takanori and {Barthelmy}, Scott D. and {Baumgartner}, Wayne H. and {Cannizzo}, John K. and {Chen}, Kevin and {Collins}, Nicholas R. and {Cummings}, Jay R. and {Gehrels}, Neil and {Krimm}, Hans A. and {Markwardt}, Craig. B. and {Palmer}, David M. and {Stamatikos}, Michael and {Troja}, Eleonora and {Ukwatta}, T.~N.},
        title = "{The Third Swift Burst Alert Telescope Gamma-Ray Burst Catalog}",
      journal = {\apj},
         year = 2016,
        month = sep,
       volume = {829},
       number = {1},
          eid = {7},
        pages = {7},
          doi = {10.3847/0004-637X/829/1/7},
archivePrefix = {arXiv},
       eprint = {1606.01956},
 primaryClass = {astro-ph.HE},
       adsurl = {https://ui.adsabs.harvard.edu/abs/2016ApJ...829....7L}
}

@ARTICLE{Liu25,
       author = {{Liu}, Y. and {Zhang}, Z.~B. and {Dong}, X.~F. and {Li}, L.~B. and {Du}, X.~Y.},
        title = "{The event rate and luminosity function of Fermi/GBM gamma-ray bursts}",
      journal = {\mnras},
         year = 2025,
        month = sep,
       volume = {542},
       number = {1},
        pages = {215-222},
          doi = {10.1093/mnras/staf1217},
archivePrefix = {arXiv},
       eprint = {2507.16595},
 primaryClass = {astro-ph.HE},
       adsurl = {https://ui.adsabs.harvard.edu/abs/2025MNRAS.542..215L}
}

@BOOK{Longair,
       author = {{Longair}, Malcolm S.},
        title = "{High Energy Astrophysics}",
         year = 2011,
        publisher = {Cambridge University Press},
       adsurl = {https://ui.adsabs.harvard.edu/abs/2011hea..book.....L}
}

@ARTICLE{Lu20,
       author = {{Lu}, Wenbin and {Bonnerot}, Cl{\'e}ment},
        title = "{Self-intersection of the fallback stream in tidal disruption events}",
      journal = {\mnras},
         year = 2020,
        month = feb,
       volume = {492},
       number = {1},
        pages = {686-707},
          doi = {10.1093/mnras/stz3405},
archivePrefix = {arXiv},
       eprint = {1904.12018},
 primaryClass = {astro-ph.HE},
       adsurl = {https://ui.adsabs.harvard.edu/abs/2020MNRAS.492..686L}
}

@ARTICLE{Luminet85,
       author = {{Luminet}, J. -P. and {Marck}, J. -A.},
        title = "{Tidal squeezing of stars by Schwarzschild black holes}",
      journal = {\mnras},
         year = 1985,
        month = jan,
       volume = {212},
        pages = {57-75},
          doi = {10.1093/mnras/212.1.57},
       adsurl = {https://ui.adsabs.harvard.edu/abs/1985MNRAS.212...57L}
}

@ARTICLE{Macleod14,
       author = {{MacLeod}, Morgan and {Goldstein}, Jacqueline and {Ramirez-Ruiz}, Enrico and {Guillochon}, James and {Samsing}, Johan},
        title = "{Illuminating Massive Black Holes with White Dwarfs: Orbital Dynamics and High-energy Transients from Tidal Interactions}",
      journal = {\apj},
         year = 2014,
        month = oct,
       volume = {794},
       number = {1},
          eid = {9},
        pages = {9},
          doi = {10.1088/0004-637X/794/1/9},
archivePrefix = {arXiv},
       eprint = {1405.1426},
 primaryClass = {astro-ph.HE},
       adsurl = {https://ui.adsabs.harvard.edu/abs/2014ApJ...794....9M}
}

@ARTICLE{Macleod16,
       author = {{MacLeod}, Morgan and {Guillochon}, James and {Ramirez-Ruiz}, Enrico and {Kasen}, Daniel and {Rosswog}, Stephan},
        title = "{Optical Thermonuclear Transients from Tidal Compression of White Dwarfs as Tracers of the Low End of the Massive Black Hole Mass Function}",
      journal = {\apj},
         year = 2016,
        month = mar,
       volume = {819},
       number = {1},
          eid = {3},
        pages = {3},
          doi = {10.3847/0004-637X/819/1/3},
archivePrefix = {arXiv},
       eprint = {1508.02399},
 primaryClass = {astro-ph.HE},
       adsurl = {https://ui.adsabs.harvard.edu/abs/2016ApJ...819....3M}
}

@ARTICLE{Maguire20,
       author = {{Maguire}, Kate and {Eracleous}, Michael and {Jonker}, Peter G. and {MacLeod}, Morgan and {Rosswog}, Stephan},
        title = "{Tidal Disruptions of White Dwarfs: Theoretical Models and Observational Prospects}",
      journal = {\ssr},
         year = 2020,
        month = mar,
       volume = {216},
       number = {3},
          eid = {39},
        pages = {39},
          doi = {10.1007/s11214-020-00661-2},
archivePrefix = {arXiv},
       eprint = {2004.00146},
 primaryClass = {astro-ph.HE},
       adsurl = {https://ui.adsabs.harvard.edu/abs/2020SSRv..216...39M}
}

@ARTICLE{Metzger12,
       author = {{Metzger}, Brian D. and {Giannios}, Dimitrios and {Mimica}, Petar},
        title = "{Afterglow model for the radio emission from the jetted tidal disruption candidate Swift J1644+57}",
      journal = {\mnras},
         year = 2012,
        month = mar,
       volume = {420},
       number = {4},
        pages = {3528-3537},
          doi = {10.1111/j.1365-2966.2011.20273.x},
archivePrefix = {arXiv},
       eprint = {1110.1111},
 primaryClass = {astro-ph.HE},
       adsurl = {https://ui.adsabs.harvard.edu/abs/2012MNRAS.420.3528M}
}

@ARTICLE{Metzger22,
       author = {{Metzger}, Brian D.},
        title = "{Cooling Envelope Model for Tidal Disruption Events}",
      journal = {\apjl},
         year = 2022,
        month = sep,
       volume = {937},
       number = {1},
          eid = {L12},
        pages = {L12},
          doi = {10.3847/2041-8213/ac90ba},
archivePrefix = {arXiv},
       eprint = {2207.07136},
 primaryClass = {astro-ph.HE},
       adsurl = {https://ui.adsabs.harvard.edu/abs/2022ApJ...937L..12M}
}

@ARTICLE{Neights25,
       author = {{Neights}, Eliza and {Burns}, Eric and {Fryer}, Chris L. and {Svinkin}, Dmitry and {Bala}, Suman and {Hamburg}, Rachel and {Gill}, Ramandeep and {Negro}, Michela and {Masterson}, Megan and {DeLaunay}, James and {Lawrence}, David J. and {Abrahams}, Sophie E.~D. and {Kawakubo}, Yuta and {Beniamini}, Paz and {Diget}, Christian Aa and {Frederiks}, Dmitry and {Goldsten}, John and {Goldstein}, Adam and {Hall-Smith}, Alexander D. and {Kara}, Erin and {Laird}, Alison M. and {Lamb}, Gavin P. and {Roberts}, Oliver J. and {Seeb}, Ryan and {Villar}, V. Ashley and {Airasca}, Aldana Holzmann and {Barber}, Joseph R. and {Bhat}, P. Narayana and {Bissaldi}, Elisabetta and {Briggs}, Michael S. and {Cleveland}, William H. and {Dalessi}, Sarah and {Depalo}, Davide and {Giles}, Misty M. and {Granot}, Jonathan and {Hristov}, Boyan A. and {Hui}, C. Michelle and {von Kienlin}, Andreas and {Kierans}, Carolyn and {Kocevski}, Daniel and {Lesage}, Stephen and {Lysenko}, Alexandra L. and {Mailyan}, Bagrat and {Malacaria}, Christian and {Mukherjee}, Oindabi and {Parsotan}, Tyler and {Ridnaia}, Anna and {Ronchini}, Samuele and {Scotton}, Lorenzo and {Trigg}, Aaron C. and {Tsvetkova}, Anastasia and {Ulanov}, Mikhail and {Veres}, P{\'e}ter and {Williams}, Maia and {Wilson-Hodge}, Colleen A. and {Wood}, Joshua},
        title = "{GRB 250702B: discovery of a gamma-ray burst from a black hole falling into a star}",
      journal = {\mnras},
         year = 2026,
        month = jan,
       volume = {545},
       number = {2},
          eid = {staf2019},
        pages = {staf2019},
          doi = {10.1093/mnras/staf2019},
       adsurl = {https://ui.adsabs.harvard.edu/abs/2026MNRAS.545f2019N}
}

@ARTICLE{Nixon21,
       author = {{Nixon}, C.~J. and {Coughlin}, Eric R. and {Miles}, Patrick R.},
        title = "{Partial, Zombie, and Full Tidal Disruption of Stars by Supermassive Black Holes}",
      journal = {\apj},
         year = 2021,
        month = dec,
       volume = {922},
       number = {2},
          eid = {168},
        pages = {168},
          doi = {10.3847/1538-4357/ac1bb8},
archivePrefix = {arXiv},
       eprint = {2108.04242},
 primaryClass = {astro-ph.HE},
       adsurl = {https://ui.adsabs.harvard.edu/abs/2021ApJ...922..168N}
}

@ARTICLE{Peters64,
       author = {{Peters}, P.~C.},
        title = "{Gravitational Radiation and the Motion of Two Point Masses}",
      journal = {Physical Review},
         year = 1964,
        month = nov,
       volume = {136},
       number = {4B},
        pages = {1224-1232},
          doi = {10.1103/PhysRev.136.B1224},
       adsurl = {https://ui.adsabs.harvard.edu/abs/1964PhRv..136.1224P}
}

%%%%%%%%%%%%%%%%%%%%%%%%%%%%%%%%%%%%%%%%%%%%%%%%%%

%%%%%%%%%%%%%%%%% APPENDICES %%%%%%%%%%%%%%%%%%%%%

%\appendix


%%%%%%%%%%%%%%%%%%%%%%%%%%%%%%%%%%%%%%%%%%%%%%%%%%

% Don't change these lines
\bsp	% typesetting comment
\label{lastpage}
\end{document}